\documentclass[a4paper,oneside,onecolumn,final,10pt]{article}

\usepackage{amsmath,amssymb,amsfonts,amsbsy,amsthm,booktabs,epsfig,graphicx,rotating,microtype}
\usepackage[left=1in,right=1in,top=1in,bottom=1in]{geometry}

\usepackage[
    backend=bibtex,
    style=numeric,
    giveninits=true,
    defernumbers=true,
    maxbibnames=99,
    url=false
]{biblatex}

\AtEveryBibitem{
    \clearfield{urlyear}
    \clearfield{urlmonth}
}

\DeclareFieldFormat*{title}{\mkbibemph{#1}}
\DeclareFieldFormat*{citetitle}{\mkbibemph{#1}}
\DeclareFieldFormat{journaltitle}{#1}

\renewbibmacro*{in:}{%
    \ifentrytype{article}
    {}
    {\printtext{\bibstring{in}\intitlepunct}}}

\newbibmacro*{pubinstorg+location+date}[1]{%
    \printlist{#1}%
    \newunit
    \printlist{location}%
    \newunit
    \usebibmacro{date}%
    \newunit}

\renewbibmacro*{publisher+location+date}{\usebibmacro{pubinstorg+location+date}{publisher}}
\renewbibmacro*{institution+location+date}{\usebibmacro{pubinstorg+location+date}{institution}}
\renewbibmacro*{organization+location+date}{\usebibmacro{pubinstorg+location+date}{organization}}

\usepackage[english]{babel}
\usepackage{hyperref}
\usepackage[capitalise,noabbrev]{cleveref}
\usepackage{glossaries}
\usepackage{graphbox}
\usepackage{listings}
\usepackage{mathtools}
\usepackage{multirow}
\usepackage{newfloat}
\usepackage{pgfplots}
\pgfplotsset{compat = 1.17}
\usepackage{siunitx}
\usepackage{soul}
\usepackage{subcaption}
\usepackage{url}

\hypersetup{
    colorlinks=true,
    linkcolor=blue,
    citecolor=blue,
    filecolor=magenta,
    urlcolor=blue,
    pdfpagemode=FullScreen,
}

\usepackage{adjustbox}
\counterwithout{figure}{section}
\counterwithout{table}{section}
\numberwithin{equation}{section}

\crefname{equation}{}{}
\crefname{figure}{Figure}{Figures}
\crefname{table}{Table}{Tables}
\crefname{algorithm}{Algorithm}{Algorithms}

\usetikzlibrary{patterns}
\usetikzlibrary{arrows}
\usetikzlibrary{shapes}
\usetikzlibrary{external}
\usetikzlibrary{arrows.meta}
\usetikzlibrary{decorations.pathreplacing}
\usetikzlibrary{positioning}

\DeclareFloatingEnvironment[
    fileext=code,
    listname={List of code listings},
    name=Code Listing,
    placement=tbhp,
    within=none,
]{code}
\crefname{code}{Code Listing}{Code Listings}

\newcommand{\inlinecpp}[1]{\lstinline[language=C++, basicstyle = \ttfamily]!#1!}

\newcommand{\vf}[1]{\vec{#1}}

\newcommand{\vv}[1]{\mathbf{#1}}

\newcommand{\ma}[1]{#1}

\newcommand{\realnumbers}{\mathbb{R}}

\newcommand{\unsigned}{\mathbb{Z}_{\geq 0}}
\newcommand{\unsignedpositive}{\mathbb{Z}_{> 0}}

\DeclarePairedDelimiterX{\norm}[1]{\lVert}{\rVert}{#1}

\newcommand{\ltwo}{L^2}

\newcommand{\femspacelagrange}{\mathbb{P}}

\newcommand{\triangulation}{\mathcal{T}}

\newcommand*\diff{\mathop{}\!\mathrm{d}}

\newacronym{amg}{AMG}{algebraic multigrid}
\newacronym{amr}{AMR}{adaptive mesh refinement}
\newacronym{aos}{AoS}{array of structures}
\newacronym{ast}{AST}{abstract syntax tree}

\newacronym{cas}{CAS}{computer algebra system}
\newacronym{cg}{CG}{conjugate gradient}
\newacronym{cgfem}{CG}{continuous Galerkin}
\newacronym{cpu}{CPU}{central processing unit}
\newacronym{crs}{CRS}{compressed row storage}

\newacronym{dg}{DG}{discontinuous Galerkin}
\newacronym[longplural=degrees of freedom]{dof}{DoF}{degree of freedom}
\newacronym{dram}{DRAM}{dynamic random access memory}

\newacronym{ecm}{ECM}{Execution-Cache-Memory}
\newacronym{eg}{EG}{enriched Galerkin}

\newacronym{fdm}{FDM}{finite difference method}
\newacronym{fem}{FEM}{finite element method}
\newacronym{fvm}{FVM}{finite volume method}
\newacronym{flops}{Flops}{floating point operations}
\newacronym{fma}{FMA}{fused multiply-add}
\newacronym{fmg}{FMG}{full multigrid}

\newacronym{gmg}{GMG}{geometric multigrid}

\newacronym{hhg}{HHG}{Hierarchical Hybrid Grids}
\newacronym{hpc}{HPC}{high performance computing}
\newacronym{hyteg}{\textsc{HyTeG}}{Hybrid Tetrahedral Grids}

\newacronym{ilu}{ILU}{incomplete LU factorization}

\newacronym{lbm}{LBM}{lattice Boltzmann method}
\newacronym[longplural=linear systems of equations, shortplural=LSEs]{lse}{LSE}{linear system of equations}

\newacronym{matvec}{MATVEC}{matrix-vector product}
\newacronym{mimd}{MIMD}{multiple instruction, multiple data}
\newacronym{minres}{MINRES}{minimum residual}
\newacronym{mpi}{MPI}{message passing interface}
\newacronym{mtbf}{MTBF}{mean time between failures}

\newacronym{pminres}{PMINRES}{preconditioned minimum residual}

\newacronym[shortplural=PDEs]{pde}{PDE}{partial differential equation}
\newacronym{pspg}{PSPG}{Pressure Stabilized Petrov-Galerkin}

\newacronym{ram}{RAM}{random access memory}

\newacronym{simd}{SIMD}{single instruction, multiple data}
\newacronym{sipg}{SIPG}{symmetric interior penalty Galerkin}
\newacronym{soa}{SoA}{structure or arrays}
\newacronym{sor}{SOR}{successive overrelaxation}
\newacronym{spd}{SPD}{symmetric and positive definite}
\newacronym{supg}{SUPG}{streamline upwind Petrov-Galerkin}

\newacronym{tme}{TME}{textbook multigrid efficiency}

\newacronym{walberla}{\textsc{waLBerla}}{widely applicable lattice Boltzmann from Erlangen}
\newacronym{wu}{WU}{work unit}

\theoremstyle{plain}%

\theoremstyle{definition}

\theoremstyle{remark}
\newtheorem{remark}{Remark}

\usepackage{authblk}

\begin{document}

    \title{Fundamental Data Structures for Matrix-Free Finite Elements on Hybrid Tetrahedral Grids}

    \author[$\ast$,$\S$]{Nils Kohl}
    \author[$\ast$,$\dagger$]{Daniel Bauer}
    \author[$\ast$,$\dagger$]{Fabian Böhm}
    \author[$\ast$,$\ddagger$]{Ulrich Rüde}
    \affil[$\ast$]{\footnotesize Friedrich-Alexander-Universität Erlangen-Nürnberg (FAU), Erlangen, Germany}
    \affil[$\dagger$]{\footnotesize Erlangen National High Performance Computing Center (NHR@FAU), Erlangen, Germany}
    \affil[$\ddagger$]{\footnotesize Centre Européen de Recherche et de Formation Avancée en Calcul Scientifique (CERFACS), Toulouse, France}
    \affil[$\S$]{Contact: Nils Kohl (\texttt{nils.kohl@fau.de})}

    \date{\vspace{-5ex}}

    \maketitle

    \begin{abstract}
    This paper presents efficient data structures for the implementation of matrix-free finite element
    methods on block-structured, hybrid tetrahedral grids.
    It provides a complete categorization of all geometric sub-objects that emerge from the regular refinement of
    the unstructured, tetrahedral coarse grid and describes efficient iteration patterns and analytical linearization
    functions for the mapping of coefficients to memory addresses.
    This foundation enables the implementation of fast, extreme-scalable, matrix-free, iterative solvers, and
    in particular geometric multigrid methods by design.
    Their application to the variable-coefficient Stokes system subject to an enriched Galerkin discretization and to
    the curl-curl problem discretized with Nédélec edge elements showcases the flexibility of the implementation.
    Eventually, the solution of a curl-curl problem with \num{1.6e11} (more than one hundred billion) unknowns on more than
    \num{32000} processes with a matrix-free full multigrid solver demonstrates its extreme-scalability.
\end{abstract}

    \section{Introduction}

The numerical approximation of \glspl*{pde} and their solutions at the extreme scale is a challenging task that
requires the design and implementation of efficient and scalable parallel algorithms and data structures.
Linear systems that arise from discretizations with very fine spatial and temporal resolution
can be subject to billions ($10^{9}$) and even trillions ($10^{12}$) of unknowns~\cite{Bauer:2020:TerraNeoMantleConvection}.
In such cases, the solution vector alone requires the allocation of multiple terabytes of main memory.
It therefore renders the explicit storage of the system matrix impossible since it typically requires of the order
of a hundred times more memory~\cite{Gmeiner:2016:QuantitativePerformanceStudy,Kohl:2022:TextbookEfficiencyMassively}.

\emph{Matrix-free} methods are crucial to solving such systems~\cite{Bergen:2006:MassivelyParallelMultigrid,
    Vos:2010:EfficientlyImplementingFinite,
    Kronbichler:2012:GenericInterfaceParallel,
    Gmeiner:2015:TextbookEfficiencyParallel,
    Gmeiner:2015:PerformanceScalabilityHierarchical,
    Kronbichler:2019:FastMatrixFreeEvaluation,
    Bastian:2019:MatrixfreeMultigridBlockpreconditioners,
    Kronbichler:2022:EnhancingDataLocality,
    Kohl:2022:TextbookEfficiencyMassively}.
Since most iterative solvers only require the result of matrix-vector operations but no explicit access
to the matrix entries, the assembly of the matrix and the evaluation of matrix-vector operations are fused and performed
on-the-fly.
On top of the reduced need for storage space, such approaches reduce the pressure on the memory bandwidth while
the arithmetic intensity of the relevant compute kernels increases, favoring the machine balance of current
cache-based architectures~\cite{Williams:2009:RooflineInsightfulVisual,Hager:2010:IntroductionHighPerformance}.

This paper introduces fundamental data structures for the implementation of massively parallel, matrix-free
finite element methods on block-structured triangular and tetrahedral grids.
The concepts herein are implemented in the \gls*{hyteg} finite element framework, building upon what has been described
in~\cite{Kohl:2019:HyTeGFiniteelementSoftware}.\footnote{\url{https://i10git.cs.fau.de/hyteg/hyteg}}

\gls*{hyteg} is a re-implementation of the \gls*{hhg} prototype
software~\cite{Bergen:2004:HierarchicalHybridGrids,Bergen:2006:HierarchicalHybridGrids}, intending to cast the
general concepts into a sustainable, extensible software framework.
With extreme-scale applications in mind, the fundamental idea of \gls*{hhg} and \gls*{hyteg} is to exploit the
block-structured grids to implement fast, matrix-free linear solvers.
Block-structured grids enable the construction of \emph{implicit} index mappings.
This is a critical advantage since it avoids the need for any bookkeeping data structures or indirect array accesses.
Due to the regular refinement algorithm applied to the coarse grid elements, a grid hierarchy that enables
straightforward implementation of geometric multigrid solvers is embedded into the framework by design.
The parallel data structures combined with solvers of optimal asymptotic complexity
using efficient matrix-free linear algebra enable the solution of systems
with trillions ($\mathcal{O}(10^{12})$) of unknowns on hundreds of thousands of parallel
processes~\cite{Gmeiner:2016:QuantitativePerformanceStudy,Kohl:2022:TextbookEfficiencyMassively}.

The \gls*{hhg} prototype implementation has successfully demonstrated the performance and extreme scalability of this
approach in a sequence of articles~\cite{Bergen:2004:HierarchicalHybridGrids,
    Bergen:2005:Times1010,
    Bergen:2006:MassivelyParallelMultigrid,
    Bergen:2006:HierarchicalHybridGrids,
    Gmeiner:2015:PerformanceScalabilityHierarchical,
    Gmeiner:2016:QuantitativePerformanceStudy,
    Bauer:2017:TwoscaleApproachEfficient,
    Bauer:2018:StencilScalingApproach,
    Bauer:2018:NewMatrixFreeApproach,
    Huber:2016:ResilienceMassivelyParallel,
    Gmeiner:2015:TextbookEfficiencyParallel,
    Waluga:2016:MasscorrectionsConservativeCoupling,
    Huber:2018:SurfaceCouplingsSubdomainWise,
    Weismuller:2015:FastAsthenosphereMotion}.
The new \gls*{hyteg} framework is a much more powerful reimplementation with more general data structures
enabling much more flexible simulations on block-structured tetrahedral grids~\cite{Kohl:2019:HyTeGFiniteelementSoftware,
    Drzisga:2019:SurrogateMatrixMethodology,
    Bauer:2020:TerraNeoMantleConvection,
    Bauer:2021:WaLBerlaBlockstructuredHighperformance,
    Kostler:2020:CodeGenerationApproaches,
    Kohl:2022:TextbookEfficiencyMassively,
    Kohl:2022:MassivelyParallelEulerianLagrangian,
    Drzisga:2022:MatrixfreeILURealization}.

This article provides a description of data structures that enable the implementation
of \emph{arbitrary} finite element discretizations for matrix-free computations on block-structured tetrahedral grids.
\Cref{sec:hyteg-domain-partitioning} introduces the underlying block-structured mesh.
\Cref{sec:hyteg-refinement} describes the regular refinement algorithm and categorizes the evolving mesh structure.
\Cref{sec:hyteg-indexing} defines an indexing scheme that supports the implementation of efficient data
structures for the implementation of matrix-free compute kernels.
\Cref{sec:hyteg-fem} discusses the design of the actual finite element spaces and matrix-free operators.
Finally, \cref{sec:hyteg-demo} demonstrates the flexibility of the concept with two illustrative example applications.
    \section{Domain partitioning}\label{sec:hyteg-domain-partitioning}

\begin{figure}[t]
    \footnotesize
    \centering
    \begin{subfigure}[t]{0.32\textwidth}
        \begin{tikzpicture}

\draw (3.62153, 3.59027) -- (3.24257, 2.71898);
\draw (3.24257, 2.71898) -- (2.99394, 3.60151);
\draw (2.99394, 3.60151) -- (3.62153, 3.59027);

\draw (3.62153, 3.59027) -- (3.24257, 2.71898);
\draw (3.24257, 2.71898) -- (4.01549, 3.10161);
\draw (4.01549, 3.10161) -- (3.62153, 3.59027);

\draw (3.62153, 3.59027) -- (3.46659, 4.15925);
\draw (3.46659, 4.15925) -- (4.15946, 3.86784);
\draw (4.15946, 3.86784) -- (3.62153, 3.59027);

\draw (3.62153, 3.59027) -- (3.46659, 4.15925);
\draw (3.46659, 4.15925) -- (2.99394, 3.60151);
\draw (2.99394, 3.60151) -- (3.62153, 3.59027);

\draw (3.62153, 3.59027) -- (4.15946, 3.86784);
\draw (4.15946, 3.86784) -- (4.01549, 3.10161);
\draw (4.01549, 3.10161) -- (3.62153, 3.59027);

\draw (5.02503, 4.95525) -- (4.14947, 4.69793);
\draw (4.14947, 4.69793) -- (4.87275, 4.11473);
\draw (4.87275, 4.11473) -- (5.02503, 4.95525);

\draw (5.02503, 4.95525) -- (4.14947, 4.69793);
\draw (4.14947, 4.69793) -- (4.45183, 5.44228);
\draw (4.45183, 5.44228) -- (5.02503, 4.95525);

\draw (5.02503, 4.95525) -- (4.87275, 4.11473);
\draw (4.87275, 4.11473) -- (5.46248, 4.49257);
\draw (5.46248, 4.49257) -- (5.02503, 4.95525);

\draw (2.34772, 5.02503) -- (2.62994, 4.15139);
\draw (2.62994, 4.15139) -- (3.20331, 4.89891);
\draw (3.20331, 4.89891) -- (2.34772, 5.02503);

\draw (2.34772, 5.02503) -- (2.62994, 4.15139);
\draw (2.62994, 4.15139) -- (1.86199, 4.45193);
\draw (1.86199, 4.45193) -- (2.34772, 5.02503);

\draw (2.34772, 5.02503) -- (3.20331, 4.89891);
\draw (3.20331, 4.89891) -- (2.8104, 5.46248);
\draw (2.8104, 5.46248) -- (2.34772, 5.02503);

\draw (4.95525, 2.27794) -- (4.73038, 3.18889);
\draw (4.73038, 3.18889) -- (4.10589, 2.44521);
\draw (4.10589, 2.44521) -- (4.95525, 2.27794);

\draw (4.95525, 2.27794) -- (4.73038, 3.18889);
\draw (4.73038, 3.18889) -- (5.44396, 2.85298);
\draw (5.44396, 2.85298) -- (4.95525, 2.27794);

\draw (4.95525, 2.27794) -- (4.10589, 2.44521);
\draw (4.10589, 2.44521) -- (4.49257, 1.84048);
\draw (4.49257, 1.84048) -- (4.95525, 2.27794);

\draw (2.27794, 2.34772) -- (3.24257, 2.71898);
\draw (3.24257, 2.71898) -- (2.42338, 3.21847);
\draw (2.42338, 3.21847) -- (2.27794, 2.34772);

\draw (2.27794, 2.34772) -- (3.24257, 2.71898);
\draw (3.24257, 2.71898) -- (2.83809, 1.86378);
\draw (2.83809, 1.86378) -- (2.27794, 2.34772);

\draw (2.27794, 2.34772) -- (2.42338, 3.21847);
\draw (2.42338, 3.21847) -- (1.84048, 2.8104);
\draw (1.84048, 2.8104) -- (2.27794, 2.34772);

\draw (3.68084, 5.74758) -- (4.14947, 4.69793);
\draw (4.14947, 4.69793) -- (3.20331, 4.89891);
\draw (3.20331, 4.89891) -- (3.68084, 5.74758);

\draw (3.68084, 5.74758) -- (4.14947, 4.69793);
\draw (4.14947, 4.69793) -- (4.45183, 5.44228);
\draw (4.45183, 5.44228) -- (3.68084, 5.74758);

\draw (3.68084, 5.74758) -- (3.20331, 4.89891);
\draw (3.20331, 4.89891) -- (2.8104, 5.46248);
\draw (2.8104, 5.46248) -- (3.68084, 5.74758);

\draw (1.55539, 3.68084) -- (2.62994, 4.15139);
\draw (2.62994, 4.15139) -- (2.42338, 3.21847);
\draw (2.42338, 3.21847) -- (1.55539, 3.68084);

\draw (1.55539, 3.68084) -- (2.62994, 4.15139);
\draw (2.62994, 4.15139) -- (1.86199, 4.45193);
\draw (1.86199, 4.45193) -- (1.55539, 3.68084);

\draw (1.55539, 3.68084) -- (2.42338, 3.21847);
\draw (2.42338, 3.21847) -- (1.84048, 2.8104);
\draw (1.84048, 2.8104) -- (1.55539, 3.68084);

\draw (5.74758, 3.62212) -- (4.73038, 3.18889);
\draw (4.73038, 3.18889) -- (4.87275, 4.11473);
\draw (4.87275, 4.11473) -- (5.74758, 3.62212);

\draw (5.74758, 3.62212) -- (4.73038, 3.18889);
\draw (4.73038, 3.18889) -- (5.44396, 2.85298);
\draw (5.44396, 2.85298) -- (5.74758, 3.62212);

\draw (5.74758, 3.62212) -- (4.87275, 4.11473);
\draw (4.87275, 4.11473) -- (5.46248, 4.49257);
\draw (5.46248, 4.49257) -- (5.74758, 3.62212);

\draw (3.62212, 1.55539) -- (3.24257, 2.71898);
\draw (3.24257, 2.71898) -- (4.10589, 2.44521);
\draw (4.10589, 2.44521) -- (3.62212, 1.55539);

\draw (3.62212, 1.55539) -- (3.24257, 2.71898);
\draw (3.24257, 2.71898) -- (2.83809, 1.86378);
\draw (2.83809, 1.86378) -- (3.62212, 1.55539);

\draw (3.62212, 1.55539) -- (4.10589, 2.44521);
\draw (4.10589, 2.44521) -- (4.49257, 1.84048);
\draw (4.49257, 1.84048) -- (3.62212, 1.55539);

\draw (4.14947, 4.69793) -- (3.20331, 4.89891);
\draw (3.20331, 4.89891) -- (3.46659, 4.15925);
\draw (3.46659, 4.15925) -- (4.14947, 4.69793);

\draw (4.14947, 4.69793) -- (4.87275, 4.11473);
\draw (4.87275, 4.11473) -- (4.15946, 3.86784);
\draw (4.15946, 3.86784) -- (4.14947, 4.69793);

\draw (4.14947, 4.69793) -- (3.46659, 4.15925);
\draw (3.46659, 4.15925) -- (4.15946, 3.86784);
\draw (4.15946, 3.86784) -- (4.14947, 4.69793);

\draw (2.62994, 4.15139) -- (3.20331, 4.89891);
\draw (3.20331, 4.89891) -- (3.46659, 4.15925);
\draw (3.46659, 4.15925) -- (2.62994, 4.15139);

\draw (2.62994, 4.15139) -- (2.42338, 3.21847);
\draw (2.42338, 3.21847) -- (2.99394, 3.60151);
\draw (2.99394, 3.60151) -- (2.62994, 4.15139);

\draw (2.62994, 4.15139) -- (3.46659, 4.15925);
\draw (3.46659, 4.15925) -- (2.99394, 3.60151);
\draw (2.99394, 3.60151) -- (2.62994, 4.15139);

\draw (4.73038, 3.18889) -- (4.87275, 4.11473);
\draw (4.87275, 4.11473) -- (4.15946, 3.86784);
\draw (4.15946, 3.86784) -- (4.73038, 3.18889);

\draw (4.73038, 3.18889) -- (4.10589, 2.44521);
\draw (4.10589, 2.44521) -- (4.01549, 3.10161);
\draw (4.01549, 3.10161) -- (4.73038, 3.18889);

\draw (4.73038, 3.18889) -- (4.15946, 3.86784);
\draw (4.15946, 3.86784) -- (4.01549, 3.10161);
\draw (4.01549, 3.10161) -- (4.73038, 3.18889);

\draw (3.24257, 2.71898) -- (2.42338, 3.21847);
\draw (2.42338, 3.21847) -- (2.99394, 3.60151);
\draw (2.99394, 3.60151) -- (3.24257, 2.71898);

\draw (3.24257, 2.71898) -- (4.10589, 2.44521);
\draw (4.10589, 2.44521) -- (4.01549, 3.10161);
\draw (4.01549, 3.10161) -- (3.24257, 2.71898);

\end{tikzpicture}
        \caption{unstructured mesh}
        \label{fig:hyteg-meshes-unstructured}
    \end{subfigure}
    \begin{subfigure}[t]{0.32\textwidth}
        \input{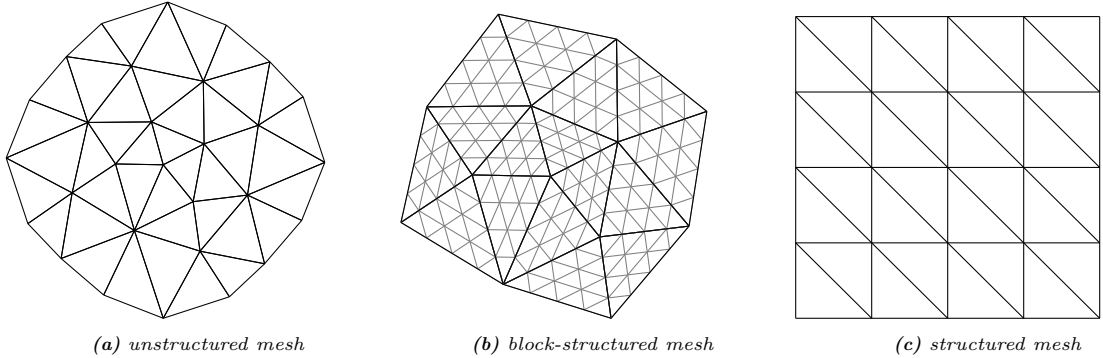}
        \caption{block-structured mesh}
        \label{fig:hyteg-meshes-block-structured}
    \end{subfigure}
    \begin{subfigure}[t]{0.32\textwidth}
        \begin{tikzpicture}

    \draw (0, 0) -- (4, 0);
    \draw (0, 1) -- (4, 1);
    \draw (0, 2) -- (4, 2);
    \draw (0, 3) -- (4, 3);
    \draw (0, 4) -- (4, 4);

    \draw (0, 0) -- (0, 4);
    \draw (1, 0) -- (1, 4);
    \draw (2, 0) -- (2, 4);
    \draw (3, 0) -- (3, 4);
    \draw (4, 0) -- (4, 4);

    \draw (1, 0) -- (0, 1);
    \draw (2, 0) -- (1, 1);
    \draw (3, 0) -- (2, 1);
    \draw (4, 0) -- (3, 1);

    \draw (1, 1) -- (0, 2);
    \draw (2, 1) -- (1, 2);
    \draw (3, 1) -- (2, 2);
    \draw (4, 1) -- (3, 2);

    \draw (1, 2) -- (0, 3);
    \draw (2, 2) -- (1, 3);
    \draw (3, 2) -- (2, 3);
    \draw (4, 2) -- (3, 3);

    \draw (1, 3) -- (0, 4);
    \draw (2, 3) -- (1, 4);
    \draw (3, 3) -- (2, 4);
    \draw (4, 3) -- (3, 4);

\end{tikzpicture}
        \caption{structured mesh}
        \label{fig:hyteg-meshes-structured}
    \end{subfigure}
    \caption{Examples of unstructured, block-structured, and structured triangular meshes.}
    \label{fig:hyteg-meshes}
\end{figure}

The mesh that approximates the domain determines the requirements for the
data structures of a simulation code.
Unstructured meshes (example in \cref{fig:hyteg-meshes-unstructured}) with varying element shapes throughout
the grid are well-suited to approximate complex domain geometries.
This comes at a cost since the data structures must incorporate the necessary bookkeeping to keep track of the
connectivity and mesh shapes over the entire computational domain.
Uniformly structured grids (example in \cref{fig:hyteg-meshes-structured}) allow for efficient iteration patterns
and data structures that can be optimized for low memory consumption since their geometry and connectivity are usually
implicitly defined.

Uniform patterns of mesh elements provide crucial optimization potential for finite element codes, e.g., since the element
integration is invariant under translation (given constant \gls*{pde} coefficients)~\cite{Bergen:2006:HierarchicalHybridGrids},
and structured iteration patterns favor the implementation of efficient compute kernels on current cache-based architectures due
to predictable memory accesses.
Block-structured meshes (example in \cref{fig:hyteg-meshes-block-structured}) are a tradeoff between
adaptability and performance.
The coarse elements provide a rough approximation of the domain.
They are then uniformly refined to exhibit a local
structure that is exploited for efficient compute kernels and data structures.
\gls*{hhg} and \gls*{hyteg} are based on this idea and employ block-structured triangular or tetrahedral elements.
Tetrahedral meshes provide more flexibility than hexahedral meshes due to their superior geometric adaptability
but require more complicated data structures.

This paper only deals with triangular and tetrahedral meshes that approximate the physical domain $\Omega$.
The triangulation $\triangulation_h(\Omega)$ of $\Omega \subset \mathbb R^d$ is defined as a partition of $\Omega$ into triangles (in two dimensions)
or tetrahedrons (in three dimensions) $T_i$.
In particular, $\triangulation_h(\Omega) \coloneqq \left\{T_1, \dots, T_{n_E}\right\}$, $T_i \subset \Omega$ open, such that
\begin{align}
    &\overline{\Omega} = \bigcup_{T_i \in \triangulation_h(\Omega)} \overline{T}_i, \label{eq:triangulation-i}\\
    &T_i \cap T_j = \emptyset \text{ for } i \neq j, \text{ and} \label{eq:triangulation-ii}\\
    &\overline{T}_i \cap \overline{T}_j,\ i \neq j \text{ is a common}
    \begin{cases}
        \text{vertex or edge }        & \text{if }d = 2, \\
        \text{vertex, edge, or face } & \text{if }d = 3. \\
    \end{cases}\label{eq:triangulation-iii}
\end{align}
The mesh size $h$ is indicated by the maximum of the diameters $h_i$ of all elements.

\begin{remark}
    Domains with curved boundaries are handled via projection of elements using suitable mapping techniques,
    e.g., those described in~\cite{Gordon:1973:TransfiniteElementMethods}.
    In \gls*{hyteg}, a surrogate approach is used to efficiently approximate arising integrals.
    Details are found in,
    e.g.,~\cite{Bauer:2017:TwoscaleApproachEfficient,Bauer:2019:LargescaleSimulationMantle,Drzisga:2019:SurrogateMatrixMethodology}.
    See \cref{sec:hyteg-demo-curl-curl} for a demonstration.
\end{remark}

\begin{figure}[t]
    \footnotesize
    \centering
    \begin{subfigure}[t]{0.32\textwidth}
        \begin{tikzpicture}

\draw (5.95078, 5.89942) -- (5.32808, 4.46773);
\draw (5.32808, 4.46773) -- (4.91955, 5.91789);
\draw (4.91955, 5.91789) -- (5.95078, 5.89942);

\draw (5.95078, 5.89942) -- (5.32808, 4.46773);
\draw (5.32808, 4.46773) -- (6.59812, 5.09646);
\draw (6.59812, 5.09646) -- (5.95078, 5.89942);

\draw (5.95078, 5.89942) -- (5.69619, 6.83435);
\draw (5.69619, 6.83435) -- (6.83469, 6.35551);
\draw (6.83469, 6.35551) -- (5.95078, 5.89942);

\draw (5.95078, 5.89942) -- (5.69619, 6.83435);
\draw (5.69619, 6.83435) -- (4.91955, 5.91789);
\draw (4.91955, 5.91789) -- (5.95078, 5.89942);

\draw (5.95078, 5.89942) -- (6.83469, 6.35551);
\draw (6.83469, 6.35551) -- (6.59812, 5.09646);
\draw (6.59812, 5.09646) -- (5.95078, 5.89942);

\draw (6.81828, 7.71949) -- (5.26358, 8.04973);
\draw (5.26358, 8.04973) -- (5.69619, 6.83435);
\draw (5.69619, 6.83435) -- (6.81828, 7.71949);

\draw (6.81828, 7.71949) -- (8.00675, 6.7612);
\draw (8.00675, 6.7612) -- (6.83469, 6.35551);
\draw (6.83469, 6.35551) -- (6.81828, 7.71949);

\draw (6.81828, 7.71949) -- (5.69619, 6.83435);
\draw (5.69619, 6.83435) -- (6.83469, 6.35551);
\draw (6.83469, 6.35551) -- (6.81828, 7.71949);

\draw (4.32143, 6.82143) -- (5.26358, 8.04973);
\draw (5.26358, 8.04973) -- (5.69619, 6.83435);
\draw (5.69619, 6.83435) -- (4.32143, 6.82143);

\draw (4.32143, 6.82143) -- (3.98201, 5.28849);
\draw (3.98201, 5.28849) -- (4.91955, 5.91789);
\draw (4.91955, 5.91789) -- (4.32143, 6.82143);

\draw (4.32143, 6.82143) -- (5.69619, 6.83435);
\draw (5.69619, 6.83435) -- (4.91955, 5.91789);
\draw (4.91955, 5.91789) -- (4.32143, 6.82143);

\draw (7.77281, 5.23988) -- (8.00675, 6.7612);
\draw (8.00675, 6.7612) -- (6.83469, 6.35551);
\draw (6.83469, 6.35551) -- (7.77281, 5.23988);

\draw (7.77281, 5.23988) -- (6.74666, 4.01789);
\draw (6.74666, 4.01789) -- (6.59812, 5.09646);
\draw (6.59812, 5.09646) -- (7.77281, 5.23988);

\draw (7.77281, 5.23988) -- (6.83469, 6.35551);
\draw (6.83469, 6.35551) -- (6.59812, 5.09646);
\draw (6.59812, 5.09646) -- (7.77281, 5.23988);

\draw (5.32808, 4.46773) -- (3.98201, 5.28849);
\draw (3.98201, 5.28849) -- (4.91955, 5.91789);
\draw (4.91955, 5.91789) -- (5.32808, 4.46773);

\draw (5.32808, 4.46773) -- (6.74666, 4.01789);
\draw (6.74666, 4.01789) -- (6.59812, 5.09646);
\draw (6.59812, 5.09646) -- (5.32808, 4.46773);
\end{tikzpicture}
        \caption{unstructured mesh}
        \label{fig:hyteg-primitive-graph-mesh}
    \end{subfigure}
    \begin{subfigure}[t]{0.32\textwidth}
        \begin{tikzpicture}

\draw (5.5373, 5.54612) -- (5.38162, 5.1882);
\draw (5.38162, 5.1882) -- (5.27949, 5.55073);
\draw (5.27949, 5.55073) -- (5.5373, 5.54612);

\draw (5.82624, 5.61309) -- (5.45262, 4.75407);
\draw (5.74453, 5.90312) -- (5.12579, 5.9142);
\draw (5.24637, 4.75777) -- (5.00125, 5.62786);

\filldraw (5.95078, 5.89942) circle (0.05);
\filldraw (5.32808, 4.46773) circle (0.05);
\filldraw (4.91955, 5.91789) circle (0.05);
\draw (5.95694, 5.34076) -- (5.80127, 4.98284);
\draw (5.80127, 4.98284) -- (6.11877, 5.14002);
\draw (6.11877, 5.14002) -- (5.95694, 5.34076);

\draw (5.82624, 5.61309) -- (5.45262, 4.75407);
\draw (6.08025, 5.73883) -- (6.46865, 5.25705);
\draw (5.58209, 4.59348) -- (6.34411, 4.97072);

\filldraw (5.95078, 5.89942) circle (0.05);
\filldraw (5.32808, 4.46773) circle (0.05);
\filldraw (6.59812, 5.09646) circle (0.05);
\draw (6.10811, 6.24718) -- (6.04446, 6.48091);
\draw (6.04446, 6.48091) -- (6.32909, 6.3612);
\draw (6.32909, 6.3612) -- (6.10811, 6.24718);

\draw (5.89986, 6.08641) -- (5.74711, 6.64736);
\draw (6.12756, 5.99064) -- (6.65791, 6.26429);
\draw (5.92389, 6.73858) -- (6.60699, 6.45128);

\filldraw (5.95078, 5.89942) circle (0.05);
\filldraw (5.69619, 6.83435) circle (0.05);
\filldraw (6.83469, 6.35551) circle (0.05);
\draw (5.62933, 6.13777) -- (5.56568, 6.3715);
\draw (5.56568, 6.3715) -- (5.37152, 6.14239);
\draw (5.37152, 6.14239) -- (5.62933, 6.13777);

\draw (5.89986, 6.08641) -- (5.74711, 6.64736);
\draw (5.74453, 5.90312) -- (5.12579, 5.9142);
\draw (5.54086, 6.65106) -- (5.07487, 6.10118);

\filldraw (5.95078, 5.89942) circle (0.05);
\filldraw (5.69619, 6.83435) circle (0.05);
\filldraw (4.91955, 5.91789) circle (0.05);
\draw (6.33359, 5.8127) -- (6.55457, 5.92673);
\draw (6.55457, 5.92673) -- (6.49543, 5.61196);
\draw (6.49543, 5.61196) -- (6.33359, 5.8127);

\draw (6.12756, 5.99064) -- (6.65791, 6.26429);
\draw (6.08025, 5.73883) -- (6.46865, 5.25705);
\draw (6.78738, 6.1037) -- (6.64543, 5.34827);

\filldraw (5.95078, 5.89942) circle (0.05);
\filldraw (6.83469, 6.35551) circle (0.05);
\filldraw (6.59812, 5.09646) circle (0.05);
\draw (6.14908, 7.58077) -- (5.76041, 7.66332);
\draw (5.76041, 7.66332) -- (5.86856, 7.35948);
\draw (5.86856, 7.35948) -- (6.14908, 7.58077);

\draw (6.50734, 7.78554) -- (5.57452, 7.98368);
\draw (6.59386, 7.54247) -- (5.92061, 7.01138);
\draw (5.3501, 7.80665) -- (5.60967, 7.07742);

\filldraw (6.81828, 7.71949) circle (0.05);
\filldraw (5.26358, 8.04973) circle (0.05);
\filldraw (5.69619, 6.83435) circle (0.05);
\draw (7.1195, 7.13892) -- (7.41662, 6.89935);
\draw (7.41662, 6.89935) -- (7.1236, 6.79793);
\draw (7.1236, 6.79793) -- (7.1195, 7.13892);

\draw (7.05597, 7.52784) -- (7.76906, 6.95286);
\draw (6.82156, 7.4467) -- (6.83141, 6.62831);
\draw (7.77234, 6.68006) -- (7.0691, 6.43665);

\filldraw (6.81828, 7.71949) circle (0.05);
\filldraw (8.00675, 6.7612) circle (0.05);
\filldraw (6.83469, 6.35551) circle (0.05);
\draw (6.54186, 7.15721) -- (6.26134, 6.93592);
\draw (6.26134, 6.93592) -- (6.54596, 6.81622);
\draw (6.54596, 6.81622) -- (6.54186, 7.15721);

\draw (6.59386, 7.54247) -- (5.92061, 7.01138);
\draw (6.82156, 7.4467) -- (6.83141, 6.62831);
\draw (5.92389, 6.73858) -- (6.60699, 6.45128);

\filldraw (6.81828, 7.71949) circle (0.05);
\filldraw (5.69619, 6.83435) circle (0.05);
\filldraw (6.83469, 6.35551) circle (0.05);
\draw (4.90066, 7.13173) -- (5.1362, 7.43881);
\draw (5.1362, 7.43881) -- (5.24435, 7.13496);
\draw (5.24435, 7.13496) -- (4.90066, 7.13173);

\draw (4.50986, 7.06709) -- (5.07515, 7.80407);
\draw (4.59638, 6.82401) -- (5.42124, 6.83176);
\draw (5.3501, 7.80665) -- (5.60967, 7.07742);

\filldraw (4.32143, 6.82143) circle (0.05);
\filldraw (5.26358, 8.04973) circle (0.05);
\filldraw (5.69619, 6.83435) circle (0.05);
\draw (4.3861, 6.21231) -- (4.30125, 5.82908);
\draw (4.30125, 5.82908) -- (4.53563, 5.98643);
\draw (4.53563, 5.98643) -- (4.3861, 6.21231);

\draw (4.25355, 6.51484) -- (4.0499, 5.59508);
\draw (4.44105, 6.64072) -- (4.79992, 6.0986);
\draw (4.16952, 5.41437) -- (4.73204, 5.79201);

\filldraw (4.32143, 6.82143) circle (0.05);
\filldraw (3.98201, 5.28849) circle (0.05);
\filldraw (4.91955, 5.91789) circle (0.05);
\draw (4.81465, 6.59877) -- (5.15834, 6.602);
\draw (5.15834, 6.602) -- (4.96418, 6.37289);
\draw (4.96418, 6.37289) -- (4.81465, 6.59877);

\draw (4.59638, 6.82401) -- (5.42124, 6.83176);
\draw (4.44105, 6.64072) -- (4.79992, 6.0986);
\draw (5.54086, 6.65106) -- (5.07487, 6.10118);

\filldraw (4.32143, 6.82143) circle (0.05);
\filldraw (5.69619, 6.83435) circle (0.05);
\filldraw (4.91955, 5.91789) circle (0.05);
\draw (7.59677, 5.89912) -- (7.65525, 6.27945);
\draw (7.65525, 6.27945) -- (7.36224, 6.17802);
\draw (7.36224, 6.17802) -- (7.59677, 5.89912);

\draw (7.8196, 5.54414) -- (7.95996, 6.45693);
\draw (7.58519, 5.463) -- (7.02232, 6.13238);
\draw (7.77234, 6.68006) -- (7.0691, 6.43665);

\filldraw (7.77281, 5.23988) circle (0.05);
\filldraw (8.00675, 6.7612) circle (0.05);
\filldraw (6.83469, 6.35551) circle (0.05);
\draw (7.2226, 4.89853) -- (6.96606, 4.59303);
\draw (6.96606, 4.59303) -- (6.92893, 4.86267);
\draw (6.92893, 4.86267) -- (7.2226, 4.89853);

\draw (7.56758, 4.99548) -- (6.95189, 4.26229);
\draw (7.53787, 5.21119) -- (6.83306, 5.12515);
\draw (6.71695, 4.23361) -- (6.62783, 4.88075);

\filldraw (7.77281, 5.23988) circle (0.05);
\filldraw (6.74666, 4.01789) circle (0.05);
\filldraw (6.59812, 5.09646) circle (0.05);
\draw (7.24461, 5.48293) -- (7.01008, 5.76184);
\draw (7.01008, 5.76184) -- (6.95093, 5.44708);
\draw (6.95093, 5.44708) -- (7.24461, 5.48293);

\draw (7.58519, 5.463) -- (7.02232, 6.13238);
\draw (7.53787, 5.21119) -- (6.83306, 5.12515);
\draw (6.78738, 6.1037) -- (6.64543, 5.34827);

\filldraw (7.77281, 5.23988) circle (0.05);
\filldraw (6.83469, 6.35551) circle (0.05);
\filldraw (6.59812, 5.09646) circle (0.05);
\draw (4.88943, 5.03546) -- (4.55291, 5.24065);
\draw (4.55291, 5.24065) -- (4.7873, 5.398);
\draw (4.7873, 5.398) -- (4.88943, 5.03546);

\draw (5.05887, 4.63189) -- (4.25123, 5.12434);
\draw (5.24637, 4.75777) -- (5.00125, 5.62786);
\draw (4.16952, 5.41437) -- (4.73204, 5.79201);

\filldraw (5.32808, 4.46773) circle (0.05);
\filldraw (3.98201, 5.28849) circle (0.05);
\filldraw (4.91955, 5.91789) circle (0.05);
\draw (6.00024, 4.51246) -- (6.35488, 4.4);
\draw (6.35488, 4.4) -- (6.31774, 4.66964);
\draw (6.31774, 4.66964) -- (6.00024, 4.51246);

\draw (5.6118, 4.37777) -- (6.46294, 4.10786);
\draw (5.58209, 4.59348) -- (6.34411, 4.97072);
\draw (6.71695, 4.23361) -- (6.62783, 4.88075);

\filldraw (5.32808, 4.46773) circle (0.05);
\filldraw (6.74666, 4.01789) circle (0.05);
\filldraw (6.59812, 5.09646) circle (0.05);

\end{tikzpicture}
        \caption{macro-primitives}
        \label{fig:hyteg-primitive-graph-primitives}
    \end{subfigure}
    \begin{subfigure}[t]{0.32\textwidth}
        \input{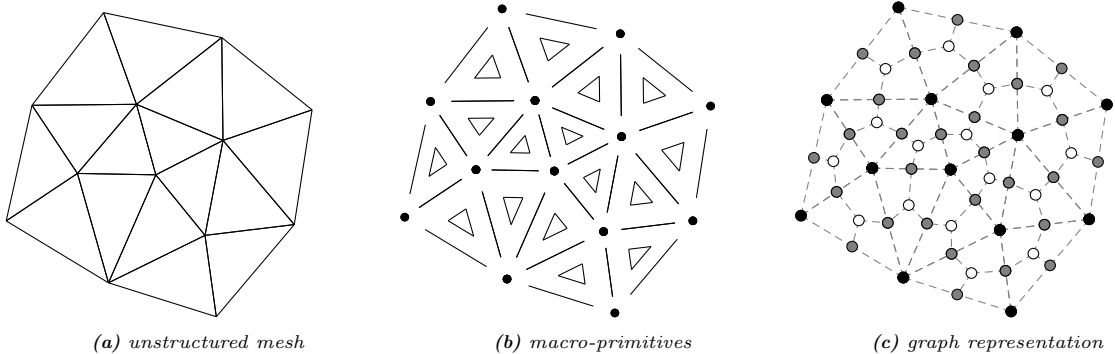}
        \caption{graph representation}
        \label{fig:hyteg-primitive-graph-graph}
    \end{subfigure}
    \caption{Unstructured mesh (\cref{fig:hyteg-primitive-graph-mesh}),
        the corresponding macro-primitives (\cref{fig:hyteg-primitive-graph-primitives}),
        and a possible graph representation (\cref{fig:hyteg-primitive-graph-graph}).
        Each macro-primitive can be interpreted as a node of a graph.
        The connectivity of the graph defines the pairs of macro-primitives that can exchange information with each other.
        It is usually discretization-dependent and can be configured as required.
        \Cref{fig:hyteg-primitive-graph-graph} illustrates a configuration with (undirected) (graph-)edges that connect
        macro-vertices (black nodes) to macro-edges (gray nodes), and macro-edges to macro-faces (white nodes).}
    \label{fig:hyteg-primitive-graph}
\end{figure}

In \gls*{hyteg} the unstructured coarse mesh is represented by an undirected graph of so-called \emph{macro-primitives}.
See \cref{fig:hyteg-primitive-graph} for a simple 2D example.
For each cell, face, edge, and vertex of the coarse mesh, a macro-cell, macro-face, macro-edge, and macro-vertex data
structure is allocated, respectively.
Macro-primitives correspond to the nodes of a graph, with (graph-)edges between neighboring primitives.
It may depend on the particular application which primitives are considered neighbors.
For instance, each macro-cell usually references the neighboring macro-faces, macro-edges, and macro-vertices.
However, for certain discretizations (e.g., \gls*{dg} schemes), references from macro-cells to neighboring macro-cells are additionally required.
Details are found in~\cite{Kohl:2019:HyTeGFiniteelementSoftware}.

On distributed parallel systems, each macro-primitive is assigned to one of the available processes.
A load balancing algorithm determines the distribution via weights for each of the graph's nodes and edges.
Each process only stores their assigned macro-primitives and metadata about those that are connected to
the local primitives.

Spatially-dependent simulation data is allocated locally on each macro-primitive
but can be migrated dynamically from one process to another one during run time.
The migration and parallel re-partitioning are vital to load balancing, coarse grid
agglomeration~\cite{Buttari:2022:BlockLowRank}, and in-memory
checkpointing~\cite{Kohl:2019:ScalableExtensibleCheckpointing}.

    \section{Refinement}\label{sec:hyteg-refinement}

This section covers the refinement process of the block-structured grid and
introduces an indexing scheme to organize the emerging geometric patterns.
All considerations herein build on the refinement algorithm due to Bey~\cite{Bey:1995:TetrahedralGridRefinement}.

\begin{remark}[Macro- and micro-primitives]
    To avoid confusion, we refer to the cells, faces, edges, and vertices that are part of the unstructured coarse mesh
    as \emph{macro}-cells, -faces, -edges, and -vertices and those created during the refinement of a macro-primitive as
    \emph{micro}-cells, -faces, -edges, and -vertices.
\end{remark}

\begin{remark}[Two dimensions]
    All following structures and indexing schemes apply to two-dimensional settings by restriction to the $xy$-plane.
\end{remark}

\subsection{Regular refinement}\label{sec:hyteg-regular-refinement}

\begin{figure}[t]
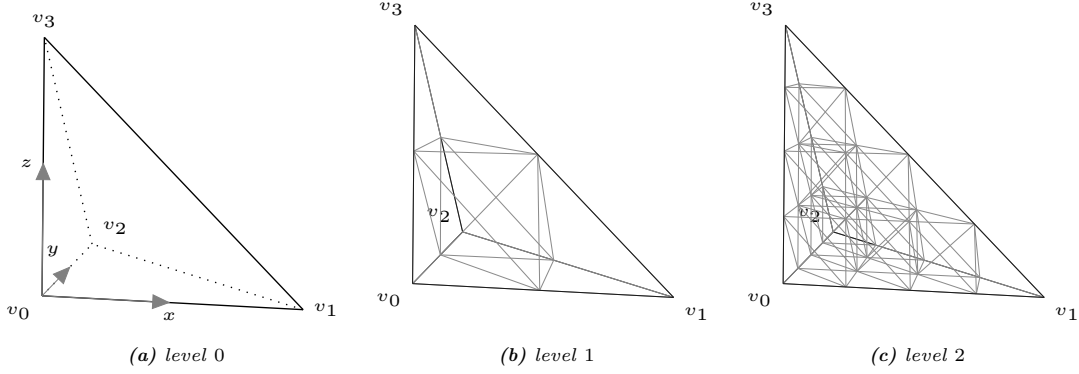

    \footnotesize
    \centering
    \begin{subfigure}[t]{.3\textwidth}
        \resizebox{\linewidth}{!}{\input{figures/macro/macro_coordinate_sytem}}
        \caption{level $0$}
        \label{fig:hyteg-tet-refinement-level-0}
    \end{subfigure}
    \begin{subfigure}[t]{.3\textwidth}
        \resizebox{\linewidth}{!}{\input{figures/macro/refinement_level_1}}
        \caption{level $1$}
    \end{subfigure}
    \begin{subfigure}[t]{.3\textwidth}
        \resizebox{\linewidth}{!}{\input{figures/macro/refinement_level_2}}
        \caption{level $2$}
    \end{subfigure}
    \caption{Uniform refinement of the reference tetrahedron according to~\cite{Bey:1995:TetrahedralGridRefinement} (i.e., \cref{eq:bey}).
    \Cref{fig:hyteg-tet-refinement-level-0} also shows the reference coordinate
    system.}
    \label{fig:hyteg-tet-refinement}
\end{figure}

The refinement process of an individual tetrahedron is described for a reference tetrahedron.
For an illustration of the structure see \cref{fig:hyteg-tet-refinement-level-0}.
Any tetrahedron from the unstructured coarse grid can be mapped to the reference tetrahedron.

To create a mesh hierarchy, the volume-primitives (macro-faces in 2D, macro-cells in 3D) are regularly refined according
to~\cite{Bey:1995:TetrahedralGridRefinement}.
Given the vertices $T = [v_0, v_1, v_2, v_3]$ of a tetrahedron, and the midpoints $v_{ij}$ of the edges connecting
$v_i$ and $v_j$, $0 \leq i,j \leq 3,\ i \neq j$, $T$ is divided into the eight subtetrahedra $T_i,\ 1 \leq i \leq 8$:
\begin{equation}\label{eq:bey}
\begin{aligned}
    T_1 &\coloneqq [\ v_{0},\  v_{01},\  v_{02},\  v_{03}\ ],\\
    T_2 &\coloneqq [\ v_{01},\  v_{1},\  v_{12},\  v_{13}\ ],\\
    T_3 &\coloneqq [\ v_{02},\  v_{12},\  v_{2},\  v_{23}\ ],\\
    T_4 &\coloneqq [\ v_{03},\  v_{13},\  v_{23},\  v_{3}\ ],\\
    T_5 &\coloneqq [\ v_{01},\  v_{02},\  v_{03},\  v_{13}\ ], \\
    T_6 &\coloneqq [\ v_{01},\  v_{02},\  v_{12},\  v_{13}\ ], \\
    T_7 &\coloneqq [\ v_{02},\  v_{03},\  v_{13},\  v_{23}\ ], \\
    T_8 &\coloneqq [\ v_{02},\  v_{12},\  v_{13},\  v_{23}\ ].
\end{aligned}
\end{equation}
This process is recursively applied to the subtetrahedra $T_1, \dots, T_8$.
The meshes of each refinement iteration are identified by levels.
Level 0 refers to the original, coarse grid tetrahedron $T$ (or the entire macro-primitive coarse grid without refinement),
level 1 to the mesh after one refinement iteration, etc.
\Cref{fig:hyteg-tet-refinement} illustrates the refinement of a single tetrahedron.
Note that this algorithm produces a globally conforming mesh, i.e., no hanging nodes are introduced at the intersection
of two coarse grid tetrahedra.

\begin{figure}[t]
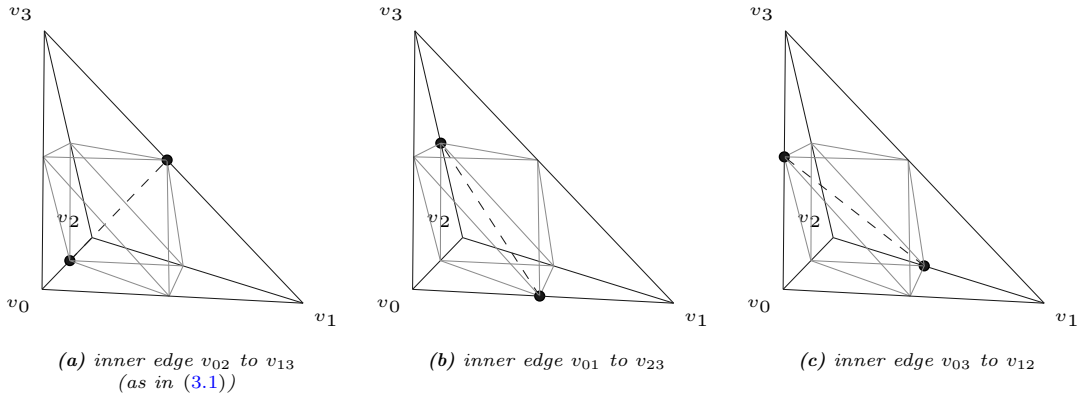

    \footnotesize
    \centering
    \begin{subfigure}[t]{.3\textwidth}
        \resizebox{\linewidth}{!}{\input{figures/macro/xyz_refinement_1_-1_1}}
        \caption{inner edge $v_{02}$ to $v_{13}$ \\ (as in \cref{eq:bey})}
    \end{subfigure}
    \begin{subfigure}[t]{.3\textwidth}
        \resizebox{\linewidth}{!}{\input{figures/macro/xyz_refinement_-1_1_1}}
        \caption{inner edge $v_{01}$ to $v_{23}$}
    \end{subfigure}
    \begin{subfigure}[t]{.3\textwidth}
        \resizebox{\linewidth}{!}{\input{figures/macro/xyz_refinement_1_1_-1}}
        \caption{inner edge $v_{03}$ to $v_{12}$}
    \end{subfigure}
    \caption{Different ways of choosing the inner edge (dashed) during the refinement process of a tetrahedron.}
    \label{fig:hyteg-macro-cell-xyz}
\end{figure}

The rule \cref{eq:bey} produces different grids depending on the permutation of the vertices $v_0, \dots, v_3$
of the original tetrahedron.
Three different possible grids may occur.
Starting from the reference tetrahedron, they are identified by the orientation of the inner edge, as illustrated in
\cref{fig:hyteg-macro-cell-xyz}.
The permutation of the local vertex indices of a macro-element enables the optimization of the refinement process
towards well-shaped elements and simplifies the implementation since the inner edge orientation can be fixed relative
to the reference tetrahedron (in \gls*{hyteg}, the inner edge always connects $v_{02}$ and
$v_{13}$ as in \cref{eq:bey}).

\subsection{Micro-primitives}\label{sec:hyteg-micro-primitives}

The recursive refinement of the reference tetrahedron, according to~\cite{Bey:1995:TetrahedralGridRefinement}, introduces
a fixed number of \emph{subgroups} of micro-primitives.
Regardless of the refinement level, each micro-primitive belongs to one of the subgroups.
Within each subgroup, micro-primitives are identical up to translation on the same refinement level and identical up to
translation and scaling on different refinement levels.

This categorization yields
\begin{itemize}
    \item 1 micro-vertex subgroup,
    \item 7 micro-edge subgroups,
    \item 12 micro-face subgroups,
    \item 6 micro-cell subgroups.
\end{itemize}

\begin{figure}[t]
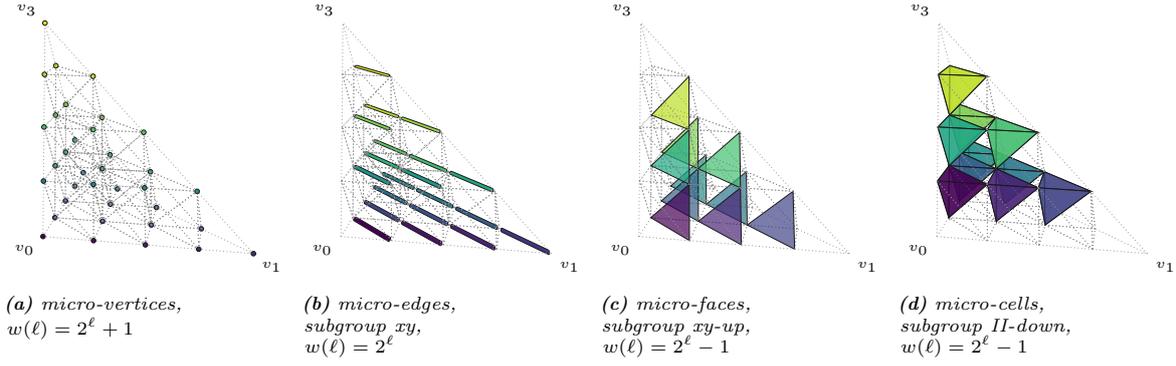

    \footnotesize
    \centering
    \begin{subfigure}[t]{.24\textwidth}
        \resizebox{\linewidth}{!}{\input{figures/micro/micro_vertices_width_5}}
        \caption{micro-vertices, \\ $w(\ell) = 2^{\ell} + 1$}
    \end{subfigure}
    \begin{subfigure}[t]{.24\textwidth}
        \resizebox{\linewidth}{!}{\input{figures/micro/micro_edge_types_width_5_cell_type_EdgeOrientation.XY}}
        \caption{micro-edges, \\ subgroup xy, \\ $w(\ell) = 2^{\ell}$}
    \end{subfigure}
    \begin{subfigure}[t]{.24\textwidth}
        \resizebox{\linewidth}{!}{\input{figures/micro/micro_face_types_width_5_cell_type_XY_UP}}
        \caption{micro-faces, \\ subgroup xy-up, \\ $w(\ell) = 2^{\ell} - 1$}
    \end{subfigure}
    \begin{subfigure}[t]{.24\textwidth}
        \resizebox{\linewidth}{!}{\input{figures/micro/micro_cell_types_width_5_cell_type_BLUE_DOWN}}
        \caption{micro-cells, \\ subgroup II-down, \\ $w(\ell) = 2^{\ell} - 1$}
    \end{subfigure}
    \caption{Illustration of one example subgroup for each micro-primitive type and the corresponding polytope side
    lengths $w(\ell)$ as functions of the refinement level $\ell \geq 2$ ($\ell = 2$ in the figures).
    Illustrations and the corresponding polytope side lengths $w(\ell)$ for all subgroups are found in \cref{sec:hyteg-all-micro-primitives},
    \cref{fig:appendix-hyteg-micro-vertices-edges,fig:appendix-hyteg-micro-faces,fig:appendix-hyteg-micro-cells,tab:appendix-hyteg-micro-primitive-subgroups}.}
    \label{fig:hyteg-micro-primitives-example}
\end{figure}

\Cref{fig:hyteg-micro-primitives-example} illustrates one subgroup of each
micro-primitive type after two refinement iterations on the reference tetrahedron.
\Cref{fig:appendix-hyteg-micro-vertices-edges,fig:appendix-hyteg-micro-faces,fig:appendix-hyteg-micro-cells} (\cref{sec:hyteg-all-micro-primitives})
give a complete overview of all subgroups.

\begin{remark}[Minumum refinement level]
    All considerations here are only valid on refinement levels $\ell \geq 2$.
    On level $\ell = 0$, for instance, there is only one micro-cell subgroup.
    Generally, on levels 0 and 1, the refined tetrahedron does not exhibit micro-primitives of all subgroups.
\end{remark}

\begin{remark}[Similar subgroups]
    The six subgroups of micro-cells are arranged into pairs, that are identical up to translation,
    reflection, and rotation.
    The naming scheme reflects this categorization (-up and -down types).
    Similar properties apply to the micro-face subgroups.
    See~\cref{fig:appendix-hyteg-micro-faces,fig:appendix-hyteg-micro-cells}.
\end{remark}

The categorization allows for a unique mapping of each element of a subgroup to the set of indices
\begin{equation}
    \label{eq:hyteg-indexing-polytope-tet}
    I_{\mathrm{tet}}(w) = \left\{ ( i, j, k ) : i + j + k < w,\ i,j,k \in \unsigned \right\},
\end{equation}
where $w \in \unsignedpositive$ is the side length of a corresponding tetrahedral polytope~\cite{Lengauer:1993:LoopParallelizationPolytope}.
The side length $w$ is equal along all macro-edges, and depends on the refinement level $\ell$.
That dependence may be different for each subgroup.
The triangular polytope is defined by
\begin{equation}
    \label{eq:hyteg-indexing-polytope-tri}
    I_{\mathrm{tri}}(w) = \left\{ ( i, j ) : i + j < w,\ i,j \in \unsigned \right\}.
\end{equation}
The cardinalities of $I_{\mathrm{tet}}(w)$ and $I_{\mathrm{tri}}(w)$ are equal to the $w^{\text{th}}$ tetrahedral number
\begin{equation}
    \label{eq:hyteg-indexing-polytope-tet-card}
    N_{\mathrm{tet}}(w) = \sum_{k=1}^{w}\left( \sum_{i=1}^{k} i \right) = \frac{w (w+1) (w+2)}{6},
\end{equation}
and the $w^{\text{th}}$ triangular number
\begin{equation}
    \label{eq:hyteg-indexing-polytope-tri-card}
    N_{\mathrm{tri}}(w) = \sum_{k=1}^{w} k = \frac{w (w+1)}{2}.
\end{equation}
The concrete values of $w$ of each subgroup illustrated in \cref{fig:hyteg-micro-primitives-example} are given in the
respective captions.
A complete list of values for all subgroups is provided in \cref{tab:appendix-hyteg-micro-primitive-subgroups}.

\begin{table}[t]
    \centering
    \footnotesize
    \begin{tabular}{llll}
        \toprule
        micro-primitive & subgroup &  $w(\ell)$ & illustration \\
        \midrule
        \multirow[t]{1}{*}{vertex} & - & $2^{\ell} + 1$ & \cref{fig:appendix-mv} \vspace{6pt}\\
        \multirow[t]{7}{*}{edge}
        & x & $2^{\ell}$ & \cref{fig:appendix-me-x} \\
        & y & $2^{\ell}$ & \cref{fig:appendix-me-y} \\
        & z & $2^{\ell}$ & \cref{fig:appendix-me-z} \\
        & xy & $2^{\ell}$ & \cref{fig:appendix-me-xy} \\
        & xz & $2^{\ell}$ & \cref{fig:appendix-me-xz} \\
        & yz & $2^{\ell}$ & \cref{fig:appendix-me-yz} \\
        & xyz & $2^{\ell} - 1$ & \cref{fig:appendix-me-xyz} \vspace{6pt} \\
        \multirow[t]{12}{*}{face}
        & z-up & $2^{\ell}$ & \cref{fig:appendix-mf-z-up} \\
        & z-down & $2^{\ell} - 1$ & \cref{fig:appendix-mf-z-down} \vspace{3pt} \\
        & y-up & $2^{\ell}$ & \cref{fig:appendix-mf-y-up} \\
        & y-down & $2^{\ell} - 1$ & \cref{fig:appendix-mf-y-down} \vspace{3pt} \\
        & x-up & $2^{\ell}$ & \cref{fig:appendix-mf-x-up} \\
        & x-down & $2^{\ell} - 1$ & \cref{fig:appendix-mf-x-down} \vspace{3pt} \\
        & xyz-up & $2^{\ell}$ & \cref{fig:appendix-mf-xyz-up} \\
        & xyz-down & $2^{\ell} - 2$ & \cref{fig:appendix-mf-xyz-down} \vspace{3pt} \\
        & xy-up & $2^{\ell} - 1$ & \cref{fig:appendix-mf-xy-up} \\
        & xy-down & $2^{\ell} - 1$ & \cref{fig:appendix-mf-xy-down} \vspace{3pt} \\
        & yz-up & $2^{\ell} - 1$ & \cref{fig:appendix-mf-yz-up} \\
        & yz-down & $2^{\ell} - 1$ & \cref{fig:appendix-mf-yz-down} \vspace{6pt} \\
        \multirow[t]{6}{*}{cell}
        & I-up & $2^{\ell}$ & \cref{fig:appendix-mc-I-up} \\
        & I-down & $2^{\ell} - 2$ & \cref{fig:appendix-mc-I-down} \vspace{3pt} \\
        & II-up & $2^{\ell} - 1$ & \cref{fig:appendix-mc-II-up} \\
        & II-down & $2^{\ell} - 1$ & \cref{fig:appendix-mc-II-down} \vspace{3pt} \\
        & III-up & $2^{\ell} - 1$ & \cref{fig:appendix-mc-III-up} \\
        & III-down & $2^{\ell} - 1$ & \cref{fig:appendix-mc-III-down} \\
        \bottomrule
    \end{tabular}
    \caption{Overview of all micro-primitive subgroups and the side lengths $w(\ell)$ of the corresponding
    tetrahedral polytopes, as functions of the refinement level $\ell \geq 2$.}
    \label{tab:appendix-hyteg-micro-primitive-subgroups}
\end{table}

The mapping of subgroups of micro-primitives to \cref{eq:hyteg-indexing-polytope-tet} facilitates the construction of
loop nests and corresponding index linearization functions, which will be discussed next, in \cref{sec:hyteg-indexing}.

    \section{Indexing}\label{sec:hyteg-indexing}

In the context of finite element discretizations, \glspl*{dof} are usually associated with the underlying mesh's vertices, edges, faces,
or volume elements.
One or more scalars (e.g., for vector-valued discretizations or when a finite element discretization associates multiple
\glspl*{dof} with a certain micro-primitive) are then allocated for each micro-primitive of a specific type.
For instance, a $\mathbb{P}_2$ finite element discretization typically associates one \gls*{dof} per vertex and one per
edge.

The coefficients are stored in arrays, and a linearization function has to be defined that uniquely maps each
\gls*{dof} to a corresponding memory address.
Due to the association of \glspl*{dof} and micro-primitives, this is accomplished via the construction of a unique
mapping of each micro-primitive to an integer.
Such a mapping can generally not be defined analytically on unstructured meshes,
and therefore requires an additional data structure.
The critical advantage of (block-) structured meshes is that they enable the construction of
\emph{implicit} index mappings and avoid the need for any bookkeeping data structures or indirect array accesses.

\begin{figure}[t]
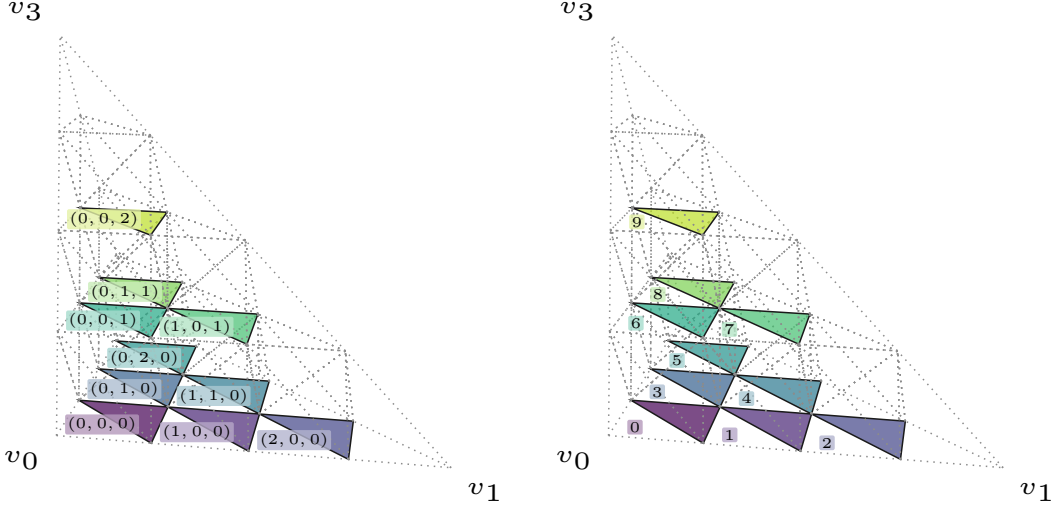

    \footnotesize
    \centering
    \begin{subfigure}[t]{.45\textwidth}
        \resizebox{\linewidth}{!}{\input{figures/micro/micro_face_types_width_5_cell_type_Z_DOWN_logical}}
    \end{subfigure}
    \begin{subfigure}[t]{.45\textwidth}
        \resizebox{\linewidth}{!}{\input{figures/micro/micro_face_types_width_5_cell_type_Z_DOWN_array}}
    \end{subfigure}
    \caption{Mapping of polytope coordinates $(i, j, k)$ (left) to consecutive integers (right) via the
    linearization function $t_w(i, j, k)$ defined in \cref{eq:hyteg-mapping-micro-primitives} for micro-faces
    of the subgroup z-down on level $\ell = 2$ ($w = 3$).}
    \label{fig:hyteg-linearization-mapping}
\end{figure}

The regular structure of each micro-primitive subgroup enables the definition of
linearization functions that only depend on the side length $w$ of the respective polytope.
A bijective map $t_w: I_{\mathrm{tet}}(w) \to \{0, \dots, N_{\mathrm{tet}}(w) - 1\}$ from the tetrahedral index set
\cref{eq:hyteg-indexing-polytope-tet} to an array index can be defined as
\begin{equation}
    \label{eq:hyteg-mapping-micro-primitives}
    \begin{aligned}
        t_w(i, j, k) \coloneqq N_{\mathrm{tet}}(w) - N_{\mathrm{tet}}(w - k)
        + N_{\mathrm{tri}}(w - k) - N_{\mathrm{tri}}(w - k - j)
        + i,
    \end{aligned}
\end{equation}
using the cardinalities \cref{eq:hyteg-indexing-polytope-tet-card,eq:hyteg-indexing-polytope-tri-card}
of the index sets \cref{eq:hyteg-indexing-polytope-tet,eq:hyteg-indexing-polytope-tri}.
\Cref{fig:hyteg-linearization-mapping} illustrates the map \cref{eq:hyteg-mapping-micro-primitives} for
the z-down micro-face subgroup.
Obviously, other mappings can be constructed.

If the number of \glspl*{dof} $m \in \unsignedpositive$ per micro-primitive is larger than $1$,
\cref{eq:hyteg-mapping-micro-primitives} cannot directly be used as a linearization function.
Two standard layouts to arrange the corresponding sequences of data sets in memory are referred to as
\gls*{aos} and \gls*{soa}.
Here, a structure is a tuple of $m$ \glspl*{dof}.

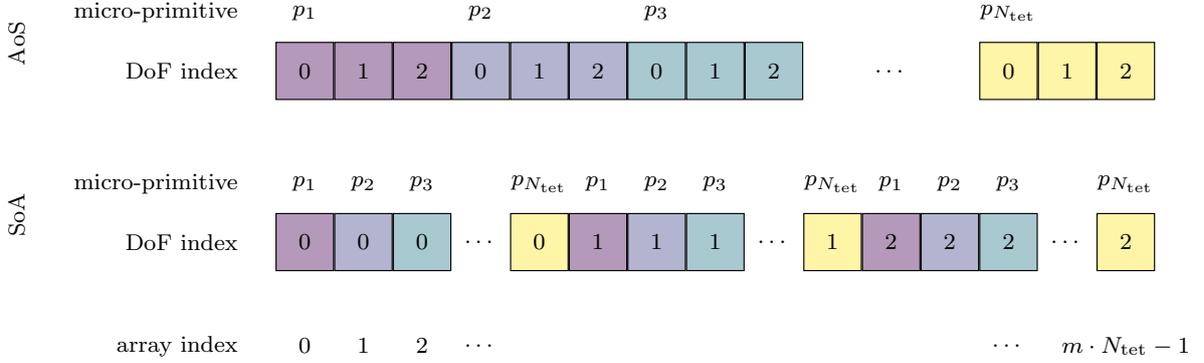
\begin{figure}[t]
    \centering
    \tiny
    \resizebox{\linewidth}{!}{%
    \begin{tikzpicture}[node distance=0pt,
        box/.style={draw, minimum width=20, minimum height=20, fill opacity=0.4, text opacity=1},
        emptybox/.style={minimum width=20, minimum height=20},
        value/.style={yshift=-1.5}]
        \footnotesize

        \definecolor{viridis_0000}{rgb}{0.267004, 0.004874, 0.329415}
        \definecolor{viridis_0200}{rgb}{0.253935, 0.265254, 0.529983}
        \definecolor{viridis_0400}{rgb}{0.163625, 0.471133, 0.558148}
        \definecolor{viridis_0600}{rgb}{0.134692, 0.658636, 0.517649}
        \definecolor{viridis_1000}{rgb}{0.993248, 0.906157, 0.143936}

        \node[box, fill=viridis_0000] (aos00)                  {$0$};
        \node[box, fill=viridis_0000] (aos01) [right=of aos00] {$1$};
        \node[box, fill=viridis_0000] (aos02) [right=of aos01] {$2$};
        \node[box, fill=viridis_0200] (aos03) [right=of aos02] {$0$};
        \node[box, fill=viridis_0200] (aos04) [right=of aos03] {$1$};
        \node[box, fill=viridis_0200] (aos05) [right=of aos04] {$2$};
        \node[box, fill=viridis_0400] (aos06) [right=of aos05] {$0$};
        \node[box, fill=viridis_0400] (aos07) [right=of aos06] {$1$};
        \node[box, fill=viridis_0400] (aos08) [right=of aos07] {$2$};
        \node[emptybox] (aos09) [right=of aos08] {};
        \node[emptybox] (aos10) [right=of aos09] {$\dots$};
        \node[emptybox] (aos11) [right=of aos10] {};
        \node[box, fill=viridis_1000] (aos12) [right=of aos11] {$0$};
        \node[box, fill=viridis_1000] (aos13) [right=of aos12] {$1$};
        \node[box, fill=viridis_1000] (aos14) [right=of aos13] {$2$};

        \node[emptybox] (aos00top) [above=of aos00] {$p_1$};
        \node[emptybox] (aos01top) [above=of aos01] {};
        \node[emptybox] (aos02top) [above=of aos02] {};
        \node[emptybox] (aos03top) [above=of aos03] {$p_2$};
        \node[emptybox] (aos04top) [above=of aos04] {};
        \node[emptybox] (aos05top) [above=of aos05] {};
        \node[emptybox] (aos06top) [above=of aos06] {$p_3$};
        \node[emptybox] (aos07top) [above=of aos07] {};
        \node[emptybox] (aos08top) [above=of aos08] {};
        \node[emptybox] (aos09top) [above=of aos09] {};
        \node[emptybox] (aos10top) [above=of aos10] {};
        \node[emptybox] (aos11top) [above=of aos11] {};
        \node[emptybox] (aos12top) [above=of aos12] {$p_{N_{\mathrm{tet}}}$};
        \node[emptybox] (aos12top) [above=of aos13] {};
        \node[emptybox] (aos12top) [above=of aos14] {};

        \node[emptybox, xshift=-90, yshift=0, rotate=90] (aosdescup) [left=of aos00top] {AoS};
        \node[emptybox, xshift=-10] (aosdescup) [left=of aos00top] {micro-primitive};
        \node[emptybox, xshift=-10] (aosdescdown) [left=of aos00] {DoF index};

        \begin{scope}[yshift=-60]
            \node[box, fill=viridis_0000] (soa00)                  {$0$};
            \node[box, fill=viridis_0200] (soa01) [right=of soa00] {$0$};
            \node[box, fill=viridis_0400] (soa02) [right=of soa01] {$0$};
            \node[emptybox] (soa03) [right=of soa02] {$\dots$};
            \node[box, fill=viridis_1000] (soa04) [right=of soa03] {$0$};
            \node[box, fill=viridis_0000] (soa05) [right=of soa04] {$1$};
            \node[box, fill=viridis_0200] (soa06) [right=of soa05] {$1$};
            \node[box, fill=viridis_0400] (soa07) [right=of soa06] {$1$};
            \node[emptybox] (soa08) [right=of soa07] {$\dots$};
            \node[box, fill=viridis_1000] (soa09) [right=of soa08] {$1$};
            \node[box, fill=viridis_0000] (soa10) [right=of soa09] {$2$};
            \node[box, fill=viridis_0200] (soa11) [right=of soa10] {$2$};
            \node[box, fill=viridis_0400] (soa12) [right=of soa11] {$2$};
            \node[emptybox] (soa13) [right=of soa12] {$\dots$};
            \node[box, fill=viridis_1000] (soa14) [right=of soa13] {$2$};

            \node[emptybox] (soa00top) [above=of soa00] {$p_1$};
            \node[emptybox] (soa01top) [above=of soa01] {$p_2$};
            \node[emptybox] (soa02top) [above=of soa02] {$p_3$};
            \node[emptybox] (soa03top) [above=of soa03] {};
            \node[emptybox] (soa04top) [above=of soa04] {$p_{N_{\mathrm{tet}}}$};
            \node[emptybox] (soa05top) [above=of soa05] {$p_1$};
            \node[emptybox] (soa06top) [above=of soa06] {$p_2$};
            \node[emptybox] (soa07top) [above=of soa07] {$p_3$};
            \node[emptybox] (soa08top) [above=of soa08] {};
            \node[emptybox] (soa09top) [above=of soa09] {$p_{N_{\mathrm{tet}}}$};
            \node[emptybox] (soa10top) [above=of soa10] {$p_1$};
            \node[emptybox] (soa11top) [above=of soa11] {$p_2$};
            \node[emptybox] (soa12top) [above=of soa12] {$p_3$};
            \node[emptybox] (soa13top) [above=of soa13] {};
            \node[emptybox] (soa14top) [above=of soa14] {$p_{N_{\mathrm{tet}}}$};

            \node[emptybox, yshift=-16] (arr00) [below=of soa00] {$0$};
            \node[emptybox, yshift=-16] (arr01) [below=of soa01] {$1$};
            \node[emptybox, yshift=-16] (arr02) [below=of soa02] {$2$};
            \node[emptybox, yshift=-16] (arr03) [below=of soa03] {$\dots$};
            \node[emptybox, yshift=-16] (arr12) [below=of soa12] {$\dots$};
            \node[emptybox, yshift=-16] (arr14) [below=of soa14] {$m \cdot N_{\mathrm{tet}} - 1$};
            \node[emptybox, xshift=-10] (arrdesc) [left=of arr00] {array index};

            \node[emptybox, xshift=-90, yshift=0, rotate=90] (soadescup) [left=of soa00top] {SoA};
            \node[emptybox, xshift=-10] (soadescup) [left=of soa00top] {micro-primitive};
            \node[emptybox, xshift=-10] (soadescdown) [left=of soa00] {DoF index};
        \end{scope}

    \end{tikzpicture}}
    \caption{Illustration of the \gls*{aos} (top) and \gls*{soa} (bottom) memory layouts for $m = 3$ \glspl*{dof}
    per micro-primitive of one subgroup.
    The micro-primitive indices are denoted by $p_\xi = (i_\xi, j_\xi, k_\xi)$. Array elements of the same color
    refer to \glspl*{dof} on the same micro-primitive, i.e., to \glspl*{dof} in the same structure.}
    \label{fig:hyteg-indexing-aos-soa}
\end{figure}

The memory layouts differ in how the structures interleave.
The \gls*{soa} layout enumerates the first element of each structure sequentially,
followed by the second element of each structure, etc., i.e.,
the first $N_{\mathrm{tet}}$ array elements correspond to the first \gls*{dof} of the $N_{\mathrm{tet}}$ micro-primitives.
In contrast, the \gls*{aos} layout linearizes each entire structure after the other, i.e.,
the first $m$ array elements correspond to the $m$ \glspl*{dof} of the first micro-primitive.
Using \cref{eq:hyteg-mapping-micro-primitives} for the linearization of the micro-primitive indices, the \gls*{aos}
and \gls*{soa} linearization functions are defined as
\begin{equation}
    \label{eq:hyteg-indexing-aos}
    t_{w, m}^{\mathrm{AoS}}(i, j, k, d) = m \cdot t_w(i, j, k) + d,
\end{equation}
and
\begin{equation}
    \label{eq:hyteg-indexing-soa}
    t_{w, m}^{\mathrm{SoA}}(i, j, k, d) = d \cdot N_{\mathrm{tet}}(w) + t_w(i, j, k),
\end{equation}
respectively, where $d \in \{0, \dots, m - 1\}$ is the index of the \gls*{dof} on the micro-primitive with index
$(i, j, k)$.
Both layouts are illustrated in \cref{fig:hyteg-indexing-aos-soa}.

The concrete selection of the iteration pattern and linearization function are crucial for the performance of
the algorithm.
Both must be designed to preserve cache locality and to enable efficient vectorization.
However, in general, the iteration pattern cannot be chosen arbitrarily since the access order modification
may influence the underlying algorithm's properties.
For instance, the convergence properties of a Gauss-Seidel iteration depend on the update
pattern~\cite{Trottenberg:2001:Multigrid}.
\Cref{code:hyteg-indexing-loop} lists two example loop nests that result in entirely consecutive access patterns for
the \gls*{aos} and \gls*{soa} linearization functions \cref{eq:hyteg-indexing-aos,eq:hyteg-indexing-soa}.
The three loops with counters $i$, $j$, and $k$ correspond to an iteration along the Cartesian coordinates $x$, $y$,
and $z$ (c.f., \cref{fig:hyteg-tet-refinement-level-0}).
The micro-primitives are traversed first in $x$-direction, then in $y$-direction, and finally in $z$-direction.
The remaining loops in \cref{code:hyteg-indexing-loop} iterate over all $m$ allocated scalars on each micro-primitive.

\begin{code}
    \centering
    \footnotesize
    \begin{lstlisting}[language=C++]
// array of structures (AoS)
for ( int k = 0; k < w; k++ )
  for ( int j = 0; j < w - k; j++ )
    for ( int i = 0; i < w - k - j; i++ )
      for ( int d = 0; d < m; d++ ) // innermost loop over structure
        int lin_idx = t_aos( w, i, j, k, d );

// structure of arrays (SoA)
for ( int d = 0; d < m; d++ )       // outermost loop over structure
  for ( int k = 0; k < w; k++ )
    for ( int j = 0; j < w - k; j++ )
      for ( int i = 0; i < w - k - j; i++ )
        int lin_idx = t_soa( w, i, j, k, d );
    \end{lstlisting}
\caption{Loop nests for the iteration over the index set $I_{\mathrm{tet}}(w)$ with $m$ \glspl*{dof} per micro-primitive
(see \cref{eq:hyteg-indexing-polytope-tet}).
Each loop nest is structured to preserve consecutive array accesses for the \gls*{aos} and \gls*{soa}
memory layouts respectively.}
\label{code:hyteg-indexing-loop}
\end{code}

    \section{Matrix-free finite elements}\label{sec:hyteg-fem}

The present section describes the construction of data structures for finite element discretizations
based on the indexing schemes introduced
in \cref{sec:hyteg-refinement,sec:hyteg-indexing} and discusses the implementation of matrix-free operators that act
on the corresponding coefficient vectors.

\subsection{Function spaces}

\begin{figure}[t]
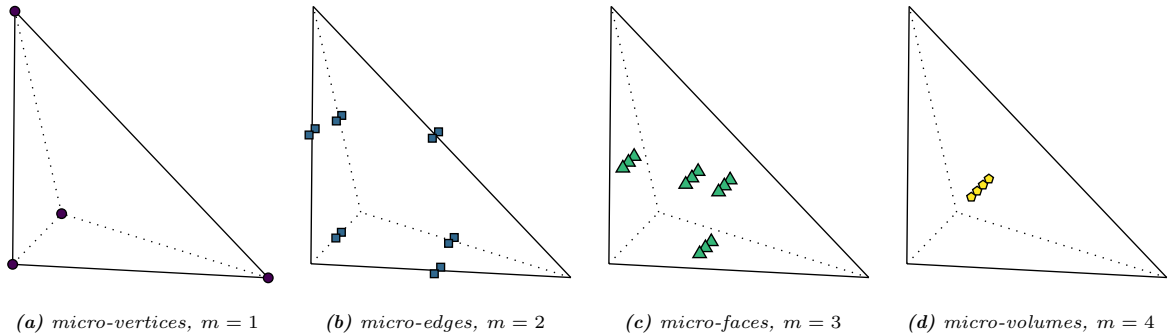

    \footnotesize
    \centering
    \begin{subfigure}[t]{.24\textwidth}
        \resizebox{\linewidth}{!}{\input{figures/fem/fem_space_v1}}
        \caption{micro-vertices, $m = 1$}
        \label{fig:hyteg-fem-elements-v}
    \end{subfigure}
    \begin{subfigure}[t]{.24\textwidth}
        \resizebox{\linewidth}{!}{\input{figures/fem/fem_space_e2}}
        \caption{micro-edges, $m = 2$}
        \label{fig:hyteg-fem-elements-e}
    \end{subfigure}
    \begin{subfigure}[t]{.24\textwidth}
        \resizebox{\linewidth}{!}{\input{figures/fem/fem_space_f3}}
        \caption{micro-faces, $m = 3$}
        \label{fig:hyteg-fem-elements-f}
    \end{subfigure}
    \begin{subfigure}[t]{.24\textwidth}
        \resizebox{\linewidth}{!}{\input{figures/fem/fem_space_c4}}
        \caption{micro-volumes, $m = 4$}
        \label{fig:hyteg-fem-elements-c}
    \end{subfigure}
    \caption{Visualization of the association of \glspl*{dof} to micro-primitives.
    The integer $m$ denotes the number of \glspl*{dof} that are associated with one micro-primitive, as discussed in \cref{sec:hyteg-indexing}.
    It is randomly chosen here for illustrative purposes.
    An overview of various types of finite element spaces and corresponding \glspl*{dof} layouts can be found in~\cite{Arnold:2014:PeriodicTableFinite}.}
    \label{fig:hyteg-fem-elements}
\end{figure}

Finite element functions can be represented by the coefficients of their basis functions.
Depending on the type of function space, one or multiple coefficients (or basis functions, respectively) are logically
associated to respective micro-primitives.
\Cref{fig:hyteg-fem-elements} illustrates some examples.

\gls*{hyteg} provides the \inlinecpp{Function} data structure for those constructions.
\mbox{\inlinecpp{Function}s} represent coefficient vectors and the corresponding finite element functions simultaneously.
Essential functionality is provided through various methods, for example, for the computation of linear combinations of
coefficient vectors, for the evaluation of the finite element function anywhere in the domain, or for the calculation of
scalar products of the coefficient vectors.

\begin{figure}[t]
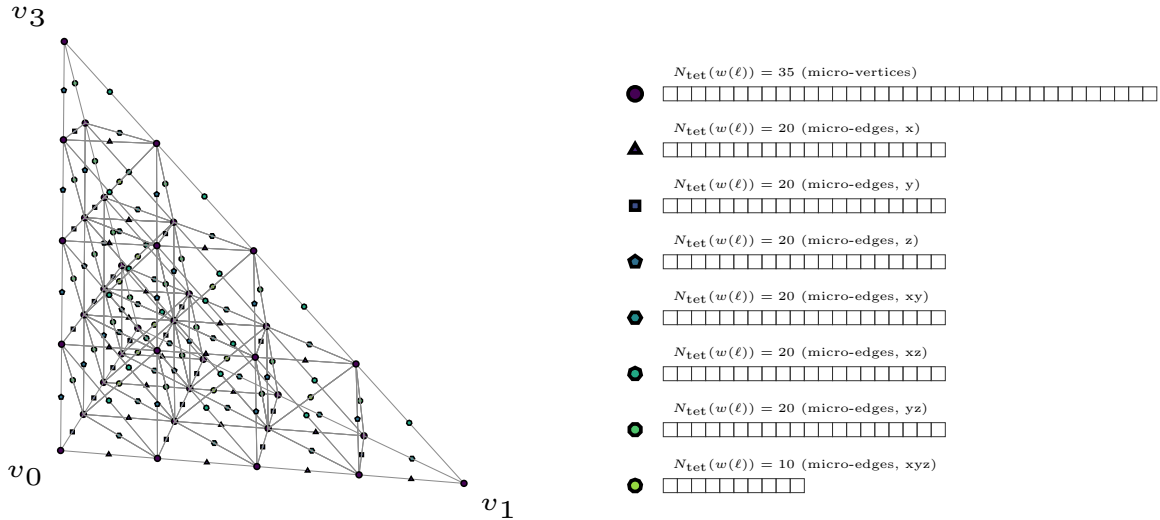

    \footnotesize
    \centering
    \begin{subfigure}[t]{.46\textwidth}
        \resizebox{\linewidth}{!}{\input{figures/fem/p2_dofs_on_refined_tet}}
        \label{fig:hyteg-fem-composition-p2-grid}
    \end{subfigure}
    \begin{subfigure}[t]{.52\textwidth}
        \raisebox{10pt}{
            \resizebox{\linewidth}{!}{\input{figures/fem/p2_dofs_on_refined_tet_arrays}}
        }
        \label{fig:hyteg-fem-composition-p2-arrays}
    \end{subfigure}
    \caption{Illustration of the composition of \glspl*{dof} of several micro-primitive subgroups into
    one function space data structure.
    In this example, one \gls*{dof} is allocated for each micro-vertex and each micro-edge, corresponding to
    a $\femspacelagrange_2$ discretization.
    The respective arrays for the individual micro-primitive subgroups are displayed on the right.
    \Glspl*{dof} that belong to the same subgroup are allocated contiguously.}
    \label{fig:hyteg-fem-composition-p2}
\end{figure}

\mbox{\inlinecpp{Function}s} can be composed to construct additional function spaces.
For instance, multiple scalar \mbox{\inlinecpp{Function}s} can be combined to represent a vector-valued function using
the \inlinecpp{VectorFunction} class.
An example composition for the $\femspacelagrange_2$ space, with vertex and edge \glspl*{dof}, is illustrated in
\cref{fig:hyteg-fem-composition-p2}.

Since \gls*{hyteg} focuses on geometric multigrid methods, the underlying coefficient vectors are usually allocated
on several levels of the mesh hierarchy.
The range of levels is specified during allocation, and virtually all methods require the user to specify the refinement
level to operate on.
This embeds the notion of a grid hierarchy into the framework and allows for the convenient construction of multigrid
methods by design.

\subsection{Operator design}\label{sec:hyteg-fem-operators}

The majority of iterative solvers and preconditioners only require the result of the action of an operator (e.g., matrix-vector multiplication) to a vector.
However, they do not need explicit access to the matrix entries.
Since the arithmetic intensity of matrix-vector operations using standard sparse matrix formats is low and does not
favor modern computer architectures, and the amount of memory required to store even sparse matrices for extreme-scale
applications is not always available, \emph{matrix-free} methods are critical to the efficient solution of extreme-scale
problems~\cite{Bergen:2006:MassivelyParallelMultigrid,
    Vos:2010:EfficientlyImplementingFinite,
    Kronbichler:2012:GenericInterfaceParallel,
    Gmeiner:2015:TextbookEfficiencyParallel,
    Gmeiner:2015:PerformanceScalabilityHierarchical,
    Kronbichler:2019:FastMatrixFreeEvaluation,
    Bastian:2019:MatrixfreeMultigridBlockpreconditioners,
    Kronbichler:2022:EnhancingDataLocality,
    Kohl:2022:TextbookEfficiencyMassively}.
The general idea is to (re-)compute the relevant matrix entries on-the-fly during the operator application instead
of precomputing and storing the entire system matrix.
This reduces storage space and bandwidth pressure since the matrix entries do not have to be loaded from memory
at the cost of additional arithmetic operations~\cite{Hager:2010:IntroductionHighPerformance}.

In \gls*{hyteg}, linear operators are therefore designed with a focus on \emph{matrix-free} computations.
The underlying block-structured grids enable crucial optimization techniques, outlined below and
studied in, e.g.,~\cite{Bergen:2004:HierarchicalHybridGrids,Gmeiner:2016:QuantitativePerformanceStudy,
    Bauer:2018:StencilScalingApproach,Kohl:2022:TextbookEfficiencyMassively}.

The general interface of all operators requires the implementation of an \inlinecpp{apply()} method that applies the
operator to a vector (i.e., to a \inlinecpp{Function} object) and writes the result to a target vector.
The \inlinecpp{apply()} method of a concrete operator in \gls*{hyteg} comprises two main ingredients:
\begin{itemize}
    \item The specification of the approximation to the bilinear form on a single element and
    \item the iteration pattern over the structured mesh on a macro-primitive.
\end{itemize}
The former defines a routine that computes local element matrices.
It is derived from the weak formulation of the underlying \gls*{pde}, the finite
element function spaces, and the order of the quadrature formula.

The latter specifies how the element matrices are applied to a vector.
While different iteration patterns are often mathematically equivalent, they must be carefully selected to achieve optimal performance.
The performance of an iteration pattern may depend on the memory layout, the type of update rule (matrix-vector multiplication,
relaxation, etc.), the function spaces, the \gls*{pde}, and the underlying hardware.
To achieve flexibility and maintainability, implementations should be composable,
such that, for instance, optimizations to iteration patterns are independent of bilinear forms.

\begin{figure}[t]
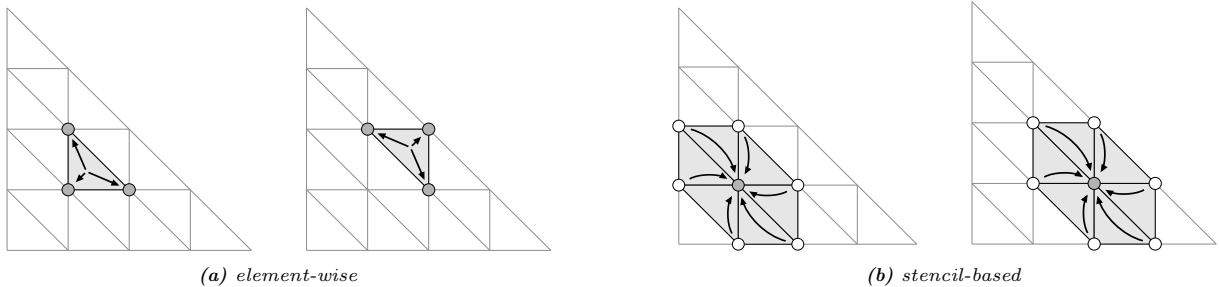

    \footnotesize
    \centering
    \begin{subfigure}[t]{.45\textwidth}
        \resizebox{.45\linewidth}{!}{\input{figures/fem/iteration_element-wise_0}}
        \hfill
        \resizebox{.45\linewidth}{!}{\input{figures/fem/iteration_element-wise_1}}
        \caption{element-wise}
        \label{fig:hyteg-fem-iteration-element-wise}
    \end{subfigure}
    \hfill
    \begin{subfigure}[t]{.45\textwidth}
        \resizebox{.45\linewidth}{!}{\input{figures/fem/iteration_stencil-based_0}}
        \hfill
        \resizebox{.45\linewidth}{!}{\input{figures/fem/iteration_stencil-based_1}}
        \caption{stencil-based}
        \label{fig:hyteg-fem-iteration-stencil-based}
    \end{subfigure}
    \caption{Illustrations of the element-wise and stencil-based update patterns.
    Each cartoon shows one update step for a discretization with vertex \glspl*{dof} as an example.
    In \cref{fig:hyteg-fem-iteration-element-wise} (element-wise pattern), two elements are traversed, with read-write access to all
    associated \glspl*{dof}. In \cref{fig:hyteg-fem-iteration-stencil-based} (stencil-based pattern) two \glspl*{dof} are traversed, with
    read-only access to \glspl*{dof} marked with empty nodes, and read-write access to the center \gls*{dof},
        marked by a filled node. The iteration order is not fixed and could be chosen differently than indicated in the
        figures. To achieve the best possible performance, it must be selected in accordance with the underlying
        memory layout.}
    \label{fig:hyteg-fem-iteration}
\end{figure}

Two typical iteration patterns for finite element discretizations are illustrated in \cref{fig:hyteg-fem-iteration}
and described in the following.

\subsubsection{Element-wise}\label{sec:hpc-matrix-free-element-wise}

Directly inferred from the standard derivation of the \gls*{fem}, \emph{element-wise} update routines iterate over the grid elements.
The weak form integrals are evaluated (or approximated) on each element to construct the local stiffness matrix.
To perform a matrix-vector multiplication, this matrix is multiplied with the \glspl*{dof} associated with the local
element, and the result is added to the destination vector.

As an example, consider the standard weak formulation of the Poisson equation.
The (global) stiffness matrix is computed via
\begin{equation}
    a_{ij} = \int_{\Omega} \nabla \phi_i \cdot \nabla \phi_j \diff x = \sum_{T \in \triangulation_h(\Omega)} \int_T \nabla \phi_i \cdot \nabla \phi_j \diff x,
\end{equation}
with basis functions $\phi_i$, and $\phi_j$.
The entry $\ma{a}^{(T)}_{kl}$ of the local stiffness matrix $\ma{A}^{(T)}$ of the element $T$ is computed as
\begin{equation}
    \ma{a}^{(T)}_{kl} = \int_T \nabla \phi_{m_T(k)} \cdot \nabla \phi_{m_T(l)} \diff x,
\end{equation}
where $k$ and $l$ are indices of \glspl*{dof} associated with the element $T$.
$m_T(\cdot)$ maps the element-local indices of the basis functions to their global indices.
The computation of the corresponding entry $\ma{a}_{ij}$ of the global system matrix may require contributions from elements other than $T$.
So instead of computing $\ma{a}_{ij}$, the result of the matrix-vector product is computed by adding the local element contributions into the destination vector:
\begin{equation}
    y_{m_T(k)} \gets y_{m_T(k)} + \sum_l \ma{a}^{(T)}_{kl} x_{m_T(l)}.
\end{equation}
\Cref{fig:hyteg-fem-iteration-element-wise} illustrates this pattern.
Obviously, $\vv{y}$ has to be appropriately initialized before the iteration.

The advantage of this approach is that operations for the integration only have to be performed once per element.
This includes, for example, calculating the transformation to the reference element.

On the downside, the actual entries $a_{ij}$ of the global system matrix are not computed explicitly.
This means that certain matrix-vector operations cannot be performed in this way.
In particular, many preconditioners require access to the diagonal entries of the system matrix.
Those usually have to be precomputed and stored in a vector.
More complicated access patterns, such as those required by Gauss-Seidel-type smoothers, are not suited
for strictly element-wise iterations in general.
Popular smoothers that only require matrix-vector products and diagonal entries include the
Chebyshev accelerated Jacobi iteration~\cite{Adams:2003:ParallelMultigridSmoothing}.
Also, element-wise patterns involve multiple store operations per vector entry when the contributions from multiple elements are summed up.
This induces more memory traffic than is theoretically necessary.

A comprehensive discussion of this element-wise approach can be found
in~\cite{Carey:1988:ElementbyelementVectorParallel,
    Kronbichler:2012:GenericInterfaceParallel,
    Kronbichler:2019:FastMatrixFreeEvaluation}, although the approach in~\cite{Carey:1988:ElementbyelementVectorParallel}
is not matrix-free (the local element matrices are precomputed and stored).

\subsubsection{Stencil-based}\label{sec:hpc-matrix-free-stencil-based}

Instead of iterating over the elements of the grid, \emph{stencil-based} update patterns iterate over the \glspl*{dof} of the destination vector.
We can interpret a stencil $S_i$ as a bipartite graph that relates an entry $y_i$ of the destination vector to those entries
$x_j$ of the source vector that correspond to the non-zero entries in row $i$ of the system matrix:
\begin{equation}
    S_i \coloneqq \{ (j, \ma{a}_{ij}) : \ma{a}_{ij} \neq 0 \}.
\end{equation}

The computation of $S_i$ requires the evaluation of integrals over all elements that are associated with $y_i$.
Using the same example as for the element-wise case, we get for each row $i$:
\begin{equation}
    \ma{a}_{ij} = \sum_{T \in \triangulation_h(\Omega)} \int_T \nabla \phi_i \cdot \nabla \phi_j \diff x, \quad \text{for all } j.
\end{equation}

The result of a matrix-vector product can then be computed for a single row $i$ by setting
\begin{equation}
    y_i \gets \sum_{j} \ma{a}_{ij} x_j.
\end{equation}
An illustration is shown in \cref{fig:hyteg-fem-iteration-stencil-based}.

The advantage of this method is that the entries of the system matrix are computed explicitly,
enabling the implementation of a variety of matrix-vector operations that are not feasible with element-wise patterns
(such as Gauss-Seidel type smoothers).
Also, each destination vector entry is written to only once.

On the downside, each update requires the evaluation of integrals on different elements, and
elements are traversed multiple times.
Although only a subset of the entries of the local stiffness matrix needs to be computed on each element, some terms
depend on the element itself and are therefore computed repeatedly.

The performance of stencil-based update patterns has been studied extensively, e.g.,
in~\cite{Bergen:2004:HierarchicalHybridGrids,Gmeiner:2015:TextbookEfficiencyParallel,Kohl:2022:TextbookEfficiencyMassively}.

Neither of the approaches is generally superior.
It depends on many factors, such as the underlying grid structure, the finite element discretization, the desired solver,
and the properties of the employed hardware.
In some instances, the element-wise and stencil-based approaches are equivalent.
Consider, for example, standard \gls*{dg} discretizations.
Since each \gls*{dof} is only associated with a single element, there is no difference between the element-wise iteration
and the stencil-based pattern.

\subsubsection{Optimizations}

Apart from reducing bandwidth pressure and memory consumption, one of the decisive advantages of matrix-free
approaches compared to pure sparse matrix computations is that they can exploit the underlying properties of the
discretization (grid structure, basis functions, etc.) to perform heavy optimizations.
An overview of standard techniques is given below, with examples of operator definitions   implemented in
\gls*{hyteg} in \cref{code:hyteg-fem-operators}.

\begin{code}[t]
    \centering
    \footnotesize
    \begin{lstlisting}[language=C++]
// Class template for a stencil-based operator
// that exploits the absence of space-dependent coefficients.
template< typename P1Form > P1ConstantOperator { ... };

// The template can be used to implement different operators via a Form object.

// Laplace operator
using P1ConstantLaplaceOperator =
    P1ConstantOperator< forms::p1_diffusion_affine_q2 >;
// Mass operator ('qe' indicates analytical integration)
using P1ConstantMassOperator = P1ConstantOperator< forms::p1_mass_affine_qe >;

// For operators with variable coefficients,
// these optimizations cannot be exploited.
// A different implementation must be used.
template< typename P1Form > P1ElementwiseOperator { ... };

// div( k(x) grad ) operator with a space-dependent coefficient k.
using P1DivKGradElementwiseOperator =
    P1ElementwiseOperator< forms::p1_div_k_grad_affine_q3 >;
    \end{lstlisting}
\caption{Examples of linear operators in \gls*{hyteg}.
The implementation is opaque to the user, and all listed operators implement the \inlinecpp{apply()} method
to perform the corresponding matrix-vector multiplication. Depending on the implementation, different
additional functionality may be supported, e.g., stencil-based operators can be employed as a Gauss-Seidel
smoother.}
\label{code:hyteg-fem-operators}
\end{code}

\paragraph{(Block-)Structured grids}

If the underlying grid is structured, matrix-free routines benefit in multiple ways.
Most importantly, the alignment of the iteration pattern and the memory layout is possible
and enables the exploitation of data locality strategies of modern cache-based processors.
In particular, this avoids scattered memory accesses in favor of contiguous accesses to the linearized data and enables vectorization.
Bandwidth pressure further reduces due to the on-the-fly computation of element vertex coordinates.
On unstructured grids, coordinates must be loaded from memory.

Eventually, the evaluation of the integrals can be optimized since the element geometry is known a priori.
The transformations from the computational elements to the reference element can be precomputed for
a small set of different elements and possibly even simplified for specific grids (axis-aligned hexahedral elements only
need to be scaled and translated).

Optimizations that exploit block-structured grids are central to the design and performance of matrix-free methods in \gls*{hhg}
and \gls*{hyteg}, and are covered in detail in, e.g.,~\cite{Bergen:2004:HierarchicalHybridGrids,
Gmeiner:2015:PerformanceScalabilityHierarchical,Kohl:2022:TextbookEfficiencyMassively,
Kohl:2019:HyTeGFiniteelementSoftware}.

\paragraph{Constant coefficients}

Additional optimizations is possible depending on the underlying \gls*{pde}.
If the operator does not involve a variable coefficient, the integration can be simplified or even computed analytically.

If both the underlying grid is structured and the operator does not involve a variable coefficient, the element matrices and
stencils for each type of element can be precomputed,
which translates to only storing one (or several) rows of the sparse system matrix on each macro-primitive,
\emph{regardless} of the refinement level.
This is arguably one of the most crucial optimizations that can be applied in matrix-free implementations and is exploited
heavily in, e.g.,~\cite{Gmeiner:2016:QuantitativePerformanceStudy,Kohl:2022:TextbookEfficiencyMassively}.

\paragraph{Surrogates}

If the grid exhibits a specific structure, but the operator involves a space-dependent coefficient, the local
element matrices and stencil entries are not constant either.
A key observation is that, for sufficiently smooth coefficients, the entries of the local stiffness matrix and the entries
of the stencil also vary smoothly over the domain.
Those coefficients can then be approximated, for instance, via polynomials.
Instead of evaluating the weak form integrals to compute a coefficient, the precomputed surrogate polynomial is evaluated.
Obviously, this introduces errors, but if implemented carefully may drastically reduce the computational effort to
obtain the entries of either the local element matrices or the stencil and still grants a sufficient approximation
of the operator.
A series of articles emerged from this idea~
\cite{Bauer:2017:TwoscaleApproachEfficient,Bauer:2018:StencilScalingApproach,Bauer:2019:LargescaleSimulationMantle,Bauer:2018:NewMatrixFreeApproach,
Drzisga:2019:SurrogateMatrixMethodology,Drzisga:2020:SurrogateMatrixMethodologyb,Drzisga:2020:SurrogateMatrixMethodologya,
Drzisga:2020:SurrogateMatrixMethodology,Drzisga:2020:StencilScalingVectorValued}.

\paragraph{Grid-aware precomputation}

Depending on the discretization and hardware, matrix-free approaches are not always superior to standard sparse-matrix implementations.
As a rule of thumb, matrix-free implementations for low-order discretizations still exhibit significant pressure on the
bandwidth, so it is unclear a priori, whether matrix-free approaches outperform standard sparse matrix
computations~\cite{Kronbichler:2018:PerformanceComparisonContinuous}.
Still, if memory \emph{consumption}, i.e., storage space, is the limiting resource, matrix-free methods are generally necessary.

However, the structured access patterns can still be exploited in the non-matrix-free case.
Instead of using a sparse format such as \gls*{crs}, the local stiffness matrices or stencils can be stored in a linearized
layout corresponding to the iteration pattern.
While this still requires loading the coefficients from memory, it avoids the frequent indirections associated with standard sparse formats.
This approach is discussed, for example,
in~\cite{Carey:1988:ElementbyelementVectorParallel,Bauer:2018:StencilScalingApproach,Tran:2022:ScalableAdaptivematrixSPMV},
and is implemented in \gls*{hyteg}.

    \section{Demonstration}\label{sec:hyteg-demo}

This section presents results on the numerical
approximation of solutions to two model problems to demonstrate the flexibility and performance
of the presented data structures in nontrivial use cases.
The performance and scalability of both demonstrator applications are subject to ongoing research
and are expected to be analyzed in forthcoming articles.
For extreme-scale studies with up to more than a trillion ($10^{12}$) \glspl*{dof} on more than
\num{140000} processes, we refer to~\cite{Kohl:2022:TextbookEfficiencyMassively,
    Kohl:2022:MassivelyParallelEulerianLagrangian}.

\begin{figure}[t]
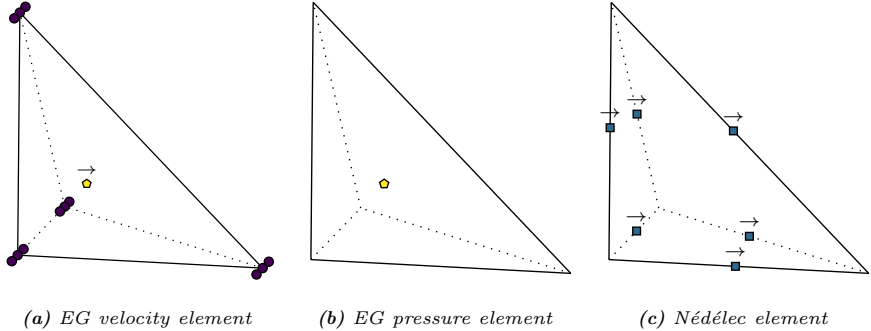

    \footnotesize
    \centering
    \begin{subfigure}[t]{.24\textwidth}
        \resizebox{\linewidth}{!}{\input{figures/fem/fem_space_v3_c1vec}}
        \caption{EG velocity element}
        \label{fig:hyteg-demo-eg-v}
    \end{subfigure}
    \begin{subfigure}[t]{.24\textwidth}
        \resizebox{\linewidth}{!}{\input{figures/fem/fem_space_c1}}
        \caption{EG pressure element}
        \label{fig:hyteg-demo-eg-p}
    \end{subfigure}
    \begin{subfigure}[t]{.24\textwidth}
        \resizebox{\linewidth}{!}{\input{figures/fem/fem_space_e1vec}}
        \caption{Nédélec element}
        \label{fig:hyteg-demo-ned}
    \end{subfigure}
    \caption{Velocity (\cref{fig:hyteg-demo-eg-v}) and pressure (\cref{fig:hyteg-demo-eg-p}) elements as employed for the
    EG discretization of the Stokes system, and Nédélec element (\cref{fig:hyteg-demo-ned}) used for the discretization
    of the curl-curl problem.
    The arrows indicate that the respective \glspl*{dof} correspond to vector-valued basis functions.}
    \label{fig:hyteg-demo-elements}
\end{figure}

\subsection{Enriched Galerkin elements for the Stokes system}

Let $\Omega \in \realnumbers^d$ be a bounded domain, $\vf{f}$ a given forcing term, and $\mu$ a
given space-dependent viscosity.
We consider the Stokes system
\begin{equation}\label{eq:hyteg-demo-stokes}
\begin{aligned}
- \nabla \cdot \left( 2 \mu \epsilon(\vf{u}) - pI\right) &= \vf{f} && \text{in}\ \Omega, \\
\nabla \cdot \vf{u} &= 0 && \text{in}\ \Omega, \\
\vf{u} &= \vf{g} && \text{on}\ \partial \Omega,
\end{aligned}
\end{equation}
where $\vf{u}$ and $p$ are the velocity and pressure solutions,
$\epsilon(\vf{u}) = \frac{1}{2}\left( \nabla \vf{u} + (\nabla \vf{u})^T \right)$
is the symmetric gradient,
and $\vf{g}$ the given velocity boundary condition.

The Stokes system \cref{eq:hyteg-demo-stokes} is discretized via the
\gls*{eg} approach described in~\cite{Yi:2022:EnrichedGalerkinMethod,Yi:2022:LockingFreeEnrichedGalerkin}.
This discretization employs a piecewise linear conforming velocity space that is enriched by a vector-valued, globally
discontinuous component:
\begin{equation}\label{eg:eg-space}
\mathbf{X}_h \coloneqq \femspacelagrange_1 \oplus \left\{ \phi \in [\ltwo(\Omega)]^d : \phi|_T = c_T(\vv{x} - \vv{x}_T),\ c_T \in \realnumbers,\ T \in \triangulation_h \right\},
\end{equation}
where $\femspacelagrange_1$ is the standard continuous Galerkin space of piecewise linear elements, and $\vv{x}_T$ is the centroid
of the element $T \in \triangulation_h$.
For the pressure, the $\femspacelagrange_0$ piecewise constant space is selected.
\Cref{fig:hyteg-demo-eg-v,fig:hyteg-demo-eg-p} illustrate the corresponding elements and \glspl*{dof}.
A single \gls*{dof} in the tetrahedron's volume represents vector-valued enrichment.
The discontinuity in the velocity (which is only in $\ltwo(\Omega)$) enforces a \gls*{dg} weak formulation and requires the
computation of jumps and averages, and edge and face integrals~\cite{Riviere:2008:DiscontinuousGalerkinMethods}.

The discretization is inf-sup stable~\cite{Yi:2022:EnrichedGalerkinMethod}.
Since the enrichment only adds a single additional \gls*{dof} per element to the piecewise linear velocity space,
it requires far fewer \glspl*{dof} per element than other inf-sup stable combinations like the traditional Taylor-Hood
pair ($\femspacelagrange_2$-$\femspacelagrange_1$)
and is thus an interesting choice for extreme-scale applications with billions or trillions of \glspl*{dof}.

\begin{figure}[t]
    \footnotesize
    \centering
    \begin{subfigure}[t]{.49\textwidth}
        \centering
        \includegraphics[width=1\textwidth]{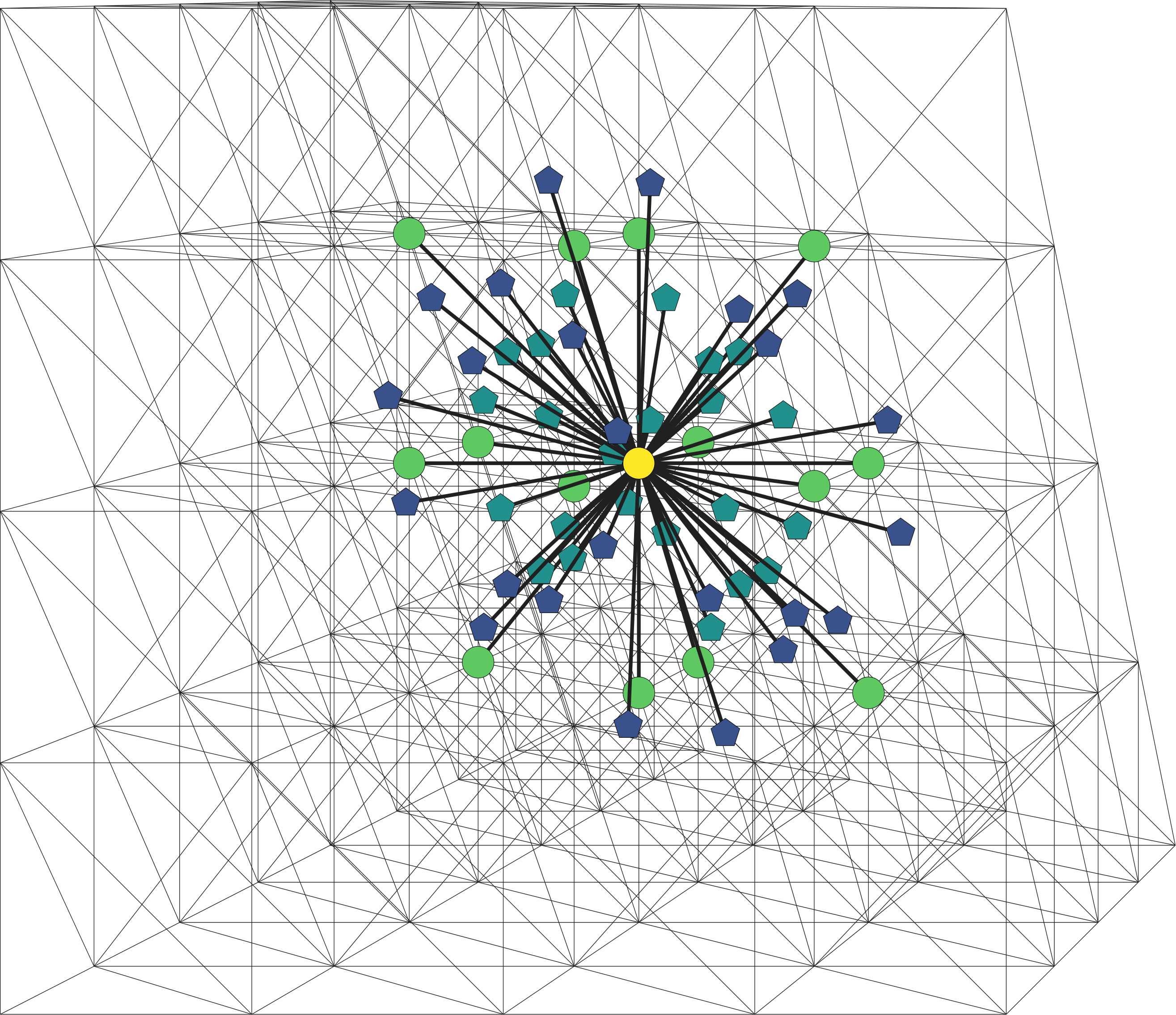}
        \caption{vertex-centered EG velocity-velocity stencil in 3D}
        \label{fig:eg-stencils-3d}
    \end{subfigure}
    \begin{subfigure}[t]{.49\textwidth}
        \centering
        \includegraphics[width=1\textwidth]{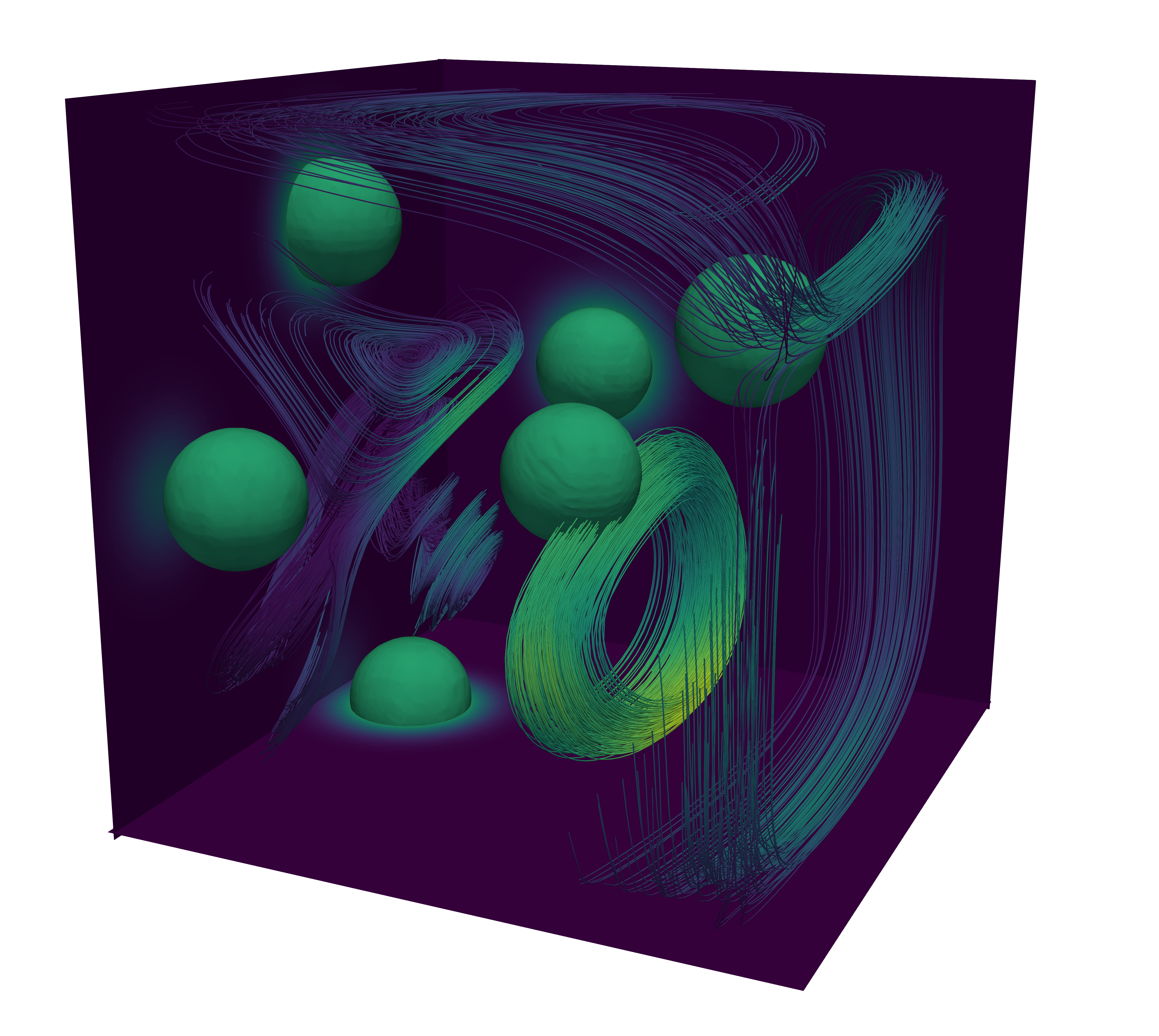}
        \caption{sinker benchmark}
        \label{fig:eg-sinker}
    \end{subfigure}
    \caption{Illustration of a 3D stencil of the \gls*{eg} discretization and a visualization of the
    computed solution of the sinker benchmark.
    \Cref{fig:eg-stencils-3d}: The vertex-centered stencil represents the coupling of the velocity components at a vertex \gls*{dof}.
    Circles correspond to vertex \glspl*{dof}; pentagons correspond to volume \glspl*{dof}. As shown in \cref{fig:hyteg-demo-eg-v},
    3 \glspl*{dof} are allocated at each micro-vertex, and a vectorial \gls*{dof} is placed in each micro-volume.
    \Cref{fig:eg-sinker}: Spherical viscosity inclusions and velocity streamlines of a computed solution of the sinker benchmark.
    Six randomly placed sinkers exhibit an exponential viscosity decay of a factor of $10^3$. A multigrid preconditioned
    MINRES solver has been applied to solve the saddle point system with roughly \num{5e5} \glspl*{dof}.
    }
    \label{fig:eg-multisinker}
\end{figure}

The \gls*{eg} discretization requires the construction of composite spaces, combining scalar vertex \glspl*{dof} and vectorial \glspl*{dof} within
the volumes for the velocity, with scalar volume \glspl*{dof} for the pressure.
The matrix-free operators map between the different spaces, e.g., from  $\femspacelagrange_1$ to the enrichment space and back.
Furthermore, the \gls*{dg} weak formulation entails integrals over lower-dimensional geometries like edges in 2D and triangles in 3D.
Those integrals also lead to dependencies across edges and faces and complicate the update patterns during the operator application.
See \cite{Bohm:2023:MatrixfreeImplementationEvaluation} for implementation details and further analysis.
\Cref{fig:eg-stencils-3d} illustrates the vertex-centered \gls*{eg} stencil that couples the velocity with itself in 3D.

We showcase the discretization for a Stokes problem with strong viscosity variations,
as frequently encountered in Earth mantle convection models~\cite{Ricard:2007:PhysicsMantleConvection,Heister:2017:HighAccuracyMantle}.
One test case that mimics this problem characteristic is the so-called \textit{multi-sinker benchmark}~\cite{May:2014:PTatin3DHighPerformanceMethods,Rudi:2017:WeightedBFBTPreconditioner}.
A variable number of spherical high-viscosity inclusions (the sinkers) are randomly placed in a 3D domain.
The viscosity rises exponentially at the boundaries of the sinkers, leading to ill-conditioned problems.
\Cref{fig:eg-multisinker} visualizes the viscosity inclusions and the velocity solution computed by \gls*{hyteg}'s \gls*{eg} implementation.

\subsection{Nédélec elements for the curl-curl problem}
\label{sec:hyteg-demo-curl-curl}

The (homogeneous) curl-curl problem
\begin{equation}\label{eq:hyteg-demo-curl-curl}
    \begin{aligned}
        \alpha\, \mathbf{curl}\, \mathbf{curl}\, \vf{u} + \beta \vf{u} &= \vf{f} && \text{in}\ \Omega, \\
        \vf{u} \times \vf{n} &= 0 && \text{on}\ \partial \Omega,
    \end{aligned}
\end{equation}
in three dimensions $(d = 3)$, with given $\vf{f}$, and $\alpha,\ \beta \in \realnumbers^+$, arises from Maxwell's
equations~\cite{Hiptmair:1998:MultigridMethodMaxwell}.
Because standard piecewise Lagrangian elements are not suited for the discretization of \cref{eq:hyteg-demo-curl-curl},
linear \emph{Nédélec} elements of the first kind~\cite{Nedelec:1980:MixedFiniteElements} are used to approximate $\vf u$ and $\vf f$.
\Cref{fig:hyteg-demo-ned} illustrates the corresponding vector-valued edge elements.
Positioning the \glspl*{dof} on the edges of the mesh enforces continuity of tangential components while components normal to cell faces are discontinuous.
This matches the amount of continuity as required for conformity in $\mathbf{H}(\mathbf{curl})$~\cite{Arnold:2018:FiniteElementExterior}.
See \cite{Bauer:2023:MultigridCurlHybrid} for implementation details.

\begin{figure}[t!]
    \footnotesize
    \null\hfill
    \begin{subfigure}[t]{.45\textwidth}
        \raisebox{30pt}{\includegraphics[width=\linewidth]{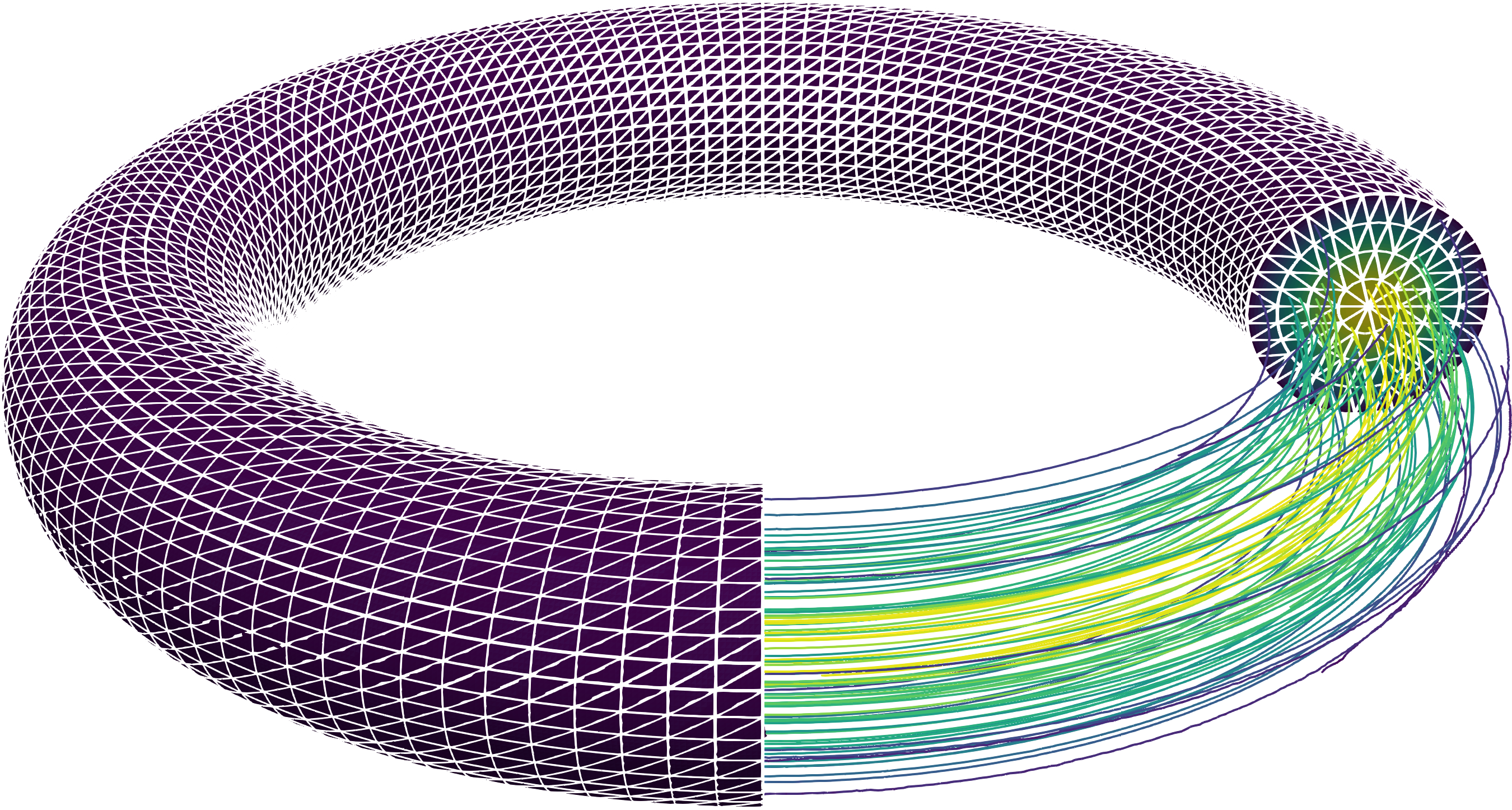}}
        \caption{coarse mesh and electric field lines of the solution on the solid torus}
        \label{fig:hyteg-demo-curl-curl-torus}
    \end{subfigure}
    \hfill
    \begin{subfigure}[t]{.5\textwidth}
        \begin{tikzpicture}
            \pgfplotsset{set layers}
            \begin{axis}[ width  = \linewidth
                        , height = 0.8\linewidth
                        , xlabel = {refinement level}
                        , ylabel = {$\|\vf e\|_{L^2}$}
                        , xtick distance   = 1
                        , ymode            = log
                        , grid             = both
                        , grid style       = {line width = .2pt, draw = gray!30}
                        , major grid style = {line width = .2pt, draw = gray!50}
                        , legend pos       = south west
                        , colormap name          = viridis
                        , cycle multiindex* list = {[colors of colormap={0, 600}]\nextlist mark list\nextlist}
                        ]
                \addplot table[y = L2error] {data/curlcurl_0256_lvl7.table.dat};
                \addlegendentry{experiment}

                \addplot[black, domain = 1:7] { 7.05098e-06 * 4^(7-x) };
                \addlegendentry{$\mathcal{O}(h^2)$}
            \end{axis}
            \begin{axis}[ width = \linewidth
                        , height = 0.8\linewidth
                        , xlabel = {\acrlongpl*{dof}}
                        , xmin = 0, xmax = 7
                        , enlarge x limits
                        , ymode = log
                        , axis x line* = top
                        , axis y line = none
                        , xtick       = {1, 4, 7}
                        , xticklabels = {\num{6.4e5}, \num{3.1e8}, \num{1.6e11}}
                        ]
                \addplot[draw = none, domain = \pgfkeysvalueof{/pgfplots/xmin}:\pgfkeysvalueof{/pgfplots/xmax}, samples = 2]{x};
            \end{axis}
        \end{tikzpicture}
        \caption{$L^2$-error grid convergence}
        \label{fig:hyteg-demo-curl-curl-convergence}
    \end{subfigure}
    \hfill\null
    \caption{Error convergence experiment using linear Nédélec elements of the first kind to solve the curl-curl problem \cref{eq:hyteg-demo-curl-curl} on the toroidal solid, discretized with \num{65280} curvilinear macro-tetrahedra.
      \Cref{eq:hyteg-demo-curl-curl} is solved using a matrix-free \gls*{fmg} solver with \num{5} V(1,1) cycles per level.
    The total number of \glspl*{dof} on the finest level is approximately \num{1.6e11}.
    }
    \label{fig:hyteg-demo-curl-curl}
\end{figure}
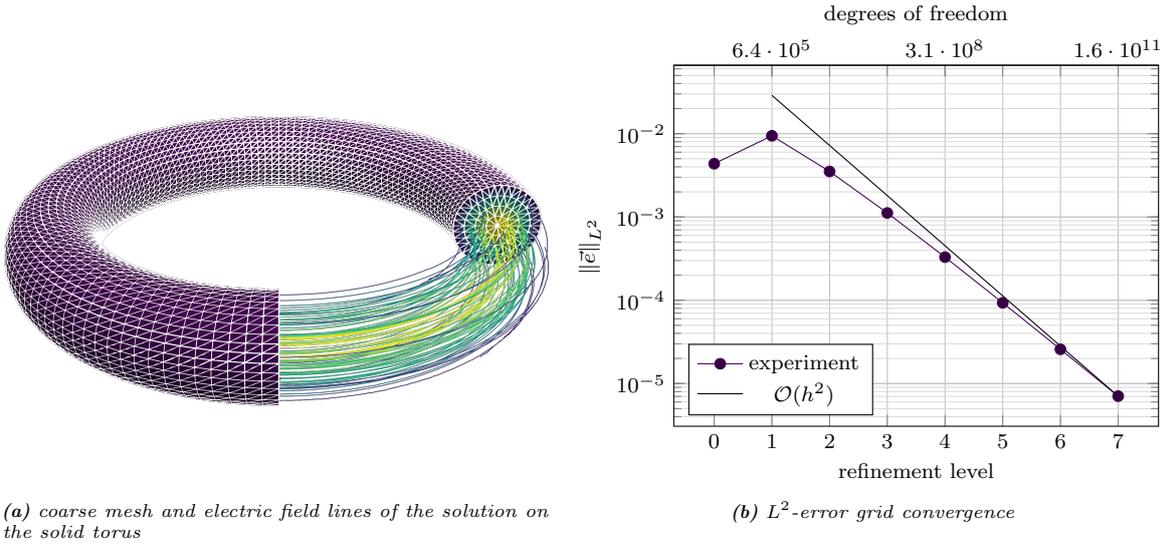

We pick the solid torus as the domain $\Omega$ and discretize it with \num{65280} macro-cells.
Because the triangulation is only a rough approximation of the curved geometry, a curvilinear mapping is
applied~\cite{Gordon:1973:TransfiniteElementMethods}.
\Cref{fig:hyteg-demo-curl-curl-torus} shows the resulting curvilinear mesh on the coarsest level.

To evaluate the grid convergence of our discretization, we construct an analytic solution with homogeneous tangential boundary conditions and determine the right-hand-side $\vf{f}$ from it.
For the sake of simplicity, we set $\alpha = \beta = 1$ in \cref{eq:hyteg-demo-curl-curl}.

Numerically, we solve the \gls*{pde} with a matrix-free \gls*{fmg} solver
performing five V(1,1) cycles on each refinement level.
Inside the V-cycles, Hiptmair's hybrid smoother~\cite{Hiptmair:1998:MultigridMethodMaxwell} is used.
Due to the non-elliptic nature of \cref{eq:hyteg-demo-curl-curl}, standard smoothers can only reduce error components orthogonal to the nullspace of the curl-operator.
Remaining error components in the nullspace of the curl-operator must be handled separately to obtain an effective multigrid scheme.
To that end, the residual remaining after smoothing is determined, lifted to the space of scalar potentials, and discretized by $\mathbb{P}_1$ elements.
Smoothing again in potential space removes the remaining oscillating error components.
The smoothed $\mathbb{P}_1$ vector must then be transformed back to the space of Nédélec elements, where it is added to the current iterate.
We choose Chebyshev smoothers of order~\num{2}~\cite{Adams:2003:ParallelMultigridSmoothing,Baker:2011:MultigridSmoothersUltraparallel} in both spaces.
This means that one hybrid smoothing step requires in total three matrix-vector products in the Nédélec space, two matrix-vector products in $\mathbb{P}_1$, and two transfer operations between the spaces.
Note that two additional $\mathbb{P}_1$ vectors must be allocated.

The system is solved up to refinement level 7, which comprises roughly
\num{1.6e11}
\glspl*{dof} on \num{32768} processes.
The numerical solution is compared against the known analytic solution and the error is measured in the $L^2$-norm.
The $L^2$-error is expected to reduce quadratically~\cite[Theorem 5.8, Remark 18]{Hiptmair:2002:FiniteElementsComputational}.
According to \cref{fig:hyteg-demo-curl-curl-convergence}, our convergence results agree with the theory.

For comparison, the same equation with jumping coefficients has recently been solved in~\cite{Kalchev:2023:ParallelElementBased} with an \gls*{amg} preconditioned \gls*{cg} solver implemented in the ParELAG miniapplication~\cite{parelag} in MFEM~\cite{Kolev:2010:ModularFiniteElement}.
The authors scaled their system up to \num{1.4e9} \glspl*{dof} on \num{4608} processes and report a total solve time of around three minutes.
Furthermore, they used a graded hexahedral mesh containing distorted elements but no curvilinear transformation.

\begin{figure}[t!]
    \footnotesize
    \null\hfill
    \begin{subfigure}[t]{.45\textwidth}
        \begin{tikzpicture}
            \begin{axis}[ width  = \linewidth
                        , height = 0.888\linewidth
                        , xlabel = {processes}
                        , ylabel = {solver runtime / s}
                        , xmode  = log
                        , ymode  = log
                        , xtick       = {64, 512, 4096, 32768}
                        , xticklabels = {\num{64}, \num{512}, \num{4096}, \num{32768}}
                        , grid             = both
                        , grid style       = {line width = .2pt, draw = gray!30}
                        , major grid style = {line width = .2pt, draw = gray!50}
                        , legend pos       = south west
                        , colormap name          = viridis
                        , cycle multiindex* list = {[colors of colormap={0, 600}]\nextlist mark list\nextlist}
                        ]
                \addplot table[x = processes, y = sumTimeMax] {data/strong-scaling.dat};
                \addlegendentry{\num{2.5e9} \acrshortpl*{dof}}

                \addplot table[x = processes, y = sumTimeMax] {data/strong-scaling-lvl4.dat};
                \addlegendentry{\num{3.1e8} \acrshortpl*{dof}}

                \addplot[black, domain = 512:32768] { 795.66637 * 512 / x };
                \addplot[black, domain =  64:32768] { 836.731   *  64 / x };
                \addlegendentry{linear scaling}
            \end{axis}
        \end{tikzpicture}
        \caption{strong scaling}
        \label{fig:hyteg-demo-curl-curl-scaling-strong}
    \end{subfigure}
    \hfill
    \begin{subfigure}[t]{.45\textwidth}
        \begin{tikzpicture}
            \pgfplotsset{set layers}
            \begin{axis}[ width  = \linewidth
                        , height = 0.888\linewidth
                        , xlabel = {processes}
                        , ylabel = {solver runtime / s}
                        , ymin   = 0
                        , xmode  = log
                        , xtick       = {64, 512, 4096, 32768}
                        , xticklabels = {\num{64}, \num{512}, \num{4096}, \num{32768}}
                        , minor y tick num = 3
                        , grid             = both
                        , grid style       = {line width = .2pt, draw = gray!30}
                        , major grid style = {line width = .2pt, draw = gray!50}
                        , colormap name          = viridis
                        , cycle multiindex* list = {[colors of colormap={0, 600}]\nextlist mark list\nextlist}
                        ]
                \addplot table[x = processes, y = sumTimeMax] {data/weak-scaling-fixed-coarse-mesh.dat};
            \end{axis}
            \begin{axis}[ width  = \linewidth
                        , height = 0.888\linewidth
                        , xlabel = {\acrlongpl*{dof}}
                        , xmin = 64, xmax = 32768
                        , enlarge x limits
                        , xmode = log
                        , axis x line* = top
                        , axis y line = none
                        , xtick       = {64, 512, 4096, 32768}
                        , xticklabels = {\num{3.1e8}, \num{2.5e9}, \num{2.0e10}, \num{1.6e11}}
                        ]
                \addplot[draw = none, domain = \pgfkeysvalueof{/pgfplots/xmin}:\pgfkeysvalueof{/pgfplots/xmax}, samples = 2]{x};
            \end{axis}
        \end{tikzpicture}
        \caption{weak scaling}
        \label{fig:hyteg-demo-curl-curl-scaling-weak}
    \end{subfigure}
    \hfill\null
    \caption{Strong and weak scalability of the \gls*{fmg} solver for the curl-curl problem \cref{eq:hyteg-demo-curl-curl}.
    \Cref{eq:hyteg-demo-curl-curl} is solved on the toroidal solid (\num{65280} curvilinear macro-tetrahedra) on different refinement levels and with varying number of processes.
    The largest run comprises roughly \num{1.6e11} \glspl*{dof} and is executed on \num{32768} cores.
    }
    \label{fig:hyteg-demo-curl-curl-scaling}
\end{figure}
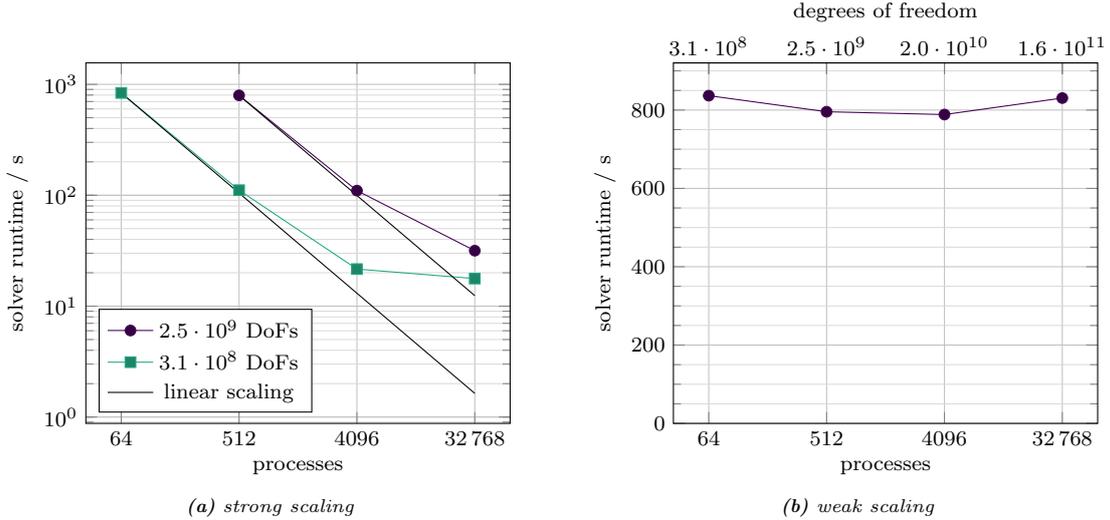

Next, we assess the strong and weak scalability of our solver.
To that end, we solve the same system on \num{32768}, \num{4096}, \num{512}, and \num{64} processes.
This corresponds to \num{256}, \num{32}, \num{4}, and \num{1} compute nodes, respectively.
The results are summarized in \cref{fig:hyteg-demo-curl-curl-scaling}.

We examine strong scalability for two problem sizes: \num{2.5e9} \glspl*{dof} (level 5) and \num{3.1e8} \glspl*{dof} (level 4).
In the larger setup, when increasing the number of processes from \num{512} to \num{32768}, a parallel efficiency of 39\% is observed.
In the smaller case, increasing the processing cores from \num{64} to \num{4096} results in a high parallel efficiency of 61\%.
Further, solving the small problem (\num{3.1e8} \glspl*{dof}) with \num{32768} processes only yields a parallel efficiency of 9\%.
This result is the expected strong scaling behavior, given that solving the system with \num{4096} processes is a matter of a few seconds.
Overall, better strong scalability is mainly hindered by the runtime on the coarser levels, where the arithmetic workload is very low compared to the amount of inter-process communication.
Note that only a few ($< 10$) \glspl*{dof} are allocated per process on the coarsest grids in both problem setups.
Solving the system directly on a higher level (i.e., using a finer coarse mesh for the \gls*{fmg} solver) might have a positive impact on the scaling results.

On the other hand, weak scaling is not impacted by this effect.
This shows in nearly constant runtimes over the range from \num{64} to \num{32768} processes.
Here, the scalability of the coarse grid solver is of less significance since 85\% of the overall runtime is spent on the finest level (level 7).

    \section{Conclusion}

This paper has provided a systematic approach to the design of data structures for arbitrary finite element
discretizations on hybrid tetrahedral grids.
It has categorized all geometric structures that emerge from regular refinement and provided
associated illustrations.
Particular focus has been put on the exploitation of the grid structure to enable efficient matrix-free kernels
and corresponding memory layouts for the coefficient vectors, regardless of the element type.
The flexibility of the implementation in the finite element framework \gls*{hyteg}
have been showcased via the variable viscosity Stokes system and the curl-curl problem.
Extreme-scalability has been demonstrated via the solution of curl-curl systems with approximately
\num{1.6e11} unknowns on more than $32000$ processes with a matrix-free full multigrid method.
The presented considerations lay the ground for the formalization, generalization, and implementation
of efficient compute kernels for large-scale, matrix-free finite element methods.

    \section*{Acknowledgments}

The authors gratefully acknowledge funding through the joint BMBF project CoMPS\footnote{\url{https://gauss-allianz.de/en/project/title/CoMPS}} (grant \texttt{16ME0647K}).
The authors would like to thank the NHR-Verein e.V.\footnote{\url{https://www.nhr-verein.de}} for supporting this work/project within the NHR Graduate School of National High Performance Computing (NHR).
The Gauss Centre for Supercomputing e.V. funded this project by providing computing time on the GCS Supercomputer HPE
Apollo Hawk at the High Performance Computing Center Stuttgart (grant \texttt{TN17/44103}).
The authors gratefully acknowledge the scientific support and HPC resources provided by the Erlangen National High
Performance Computing Center (NHR@FAU) of the Friedrich-Alexander-Universität Erlangen-Nürnberg (FAU).
NHR funding is provided by federal and Bavarian state authorities.
NHR@FAU hardware is partially funded by the German Research Foundation (DFG) – 440719683.

    \printbibliography

    \newpage
\appendix
\section{Illustrations of all micro-primitive subgroups}\label{sec:hyteg-all-micro-primitives}

\vspace*{\fill}

\begin{figure}[h]
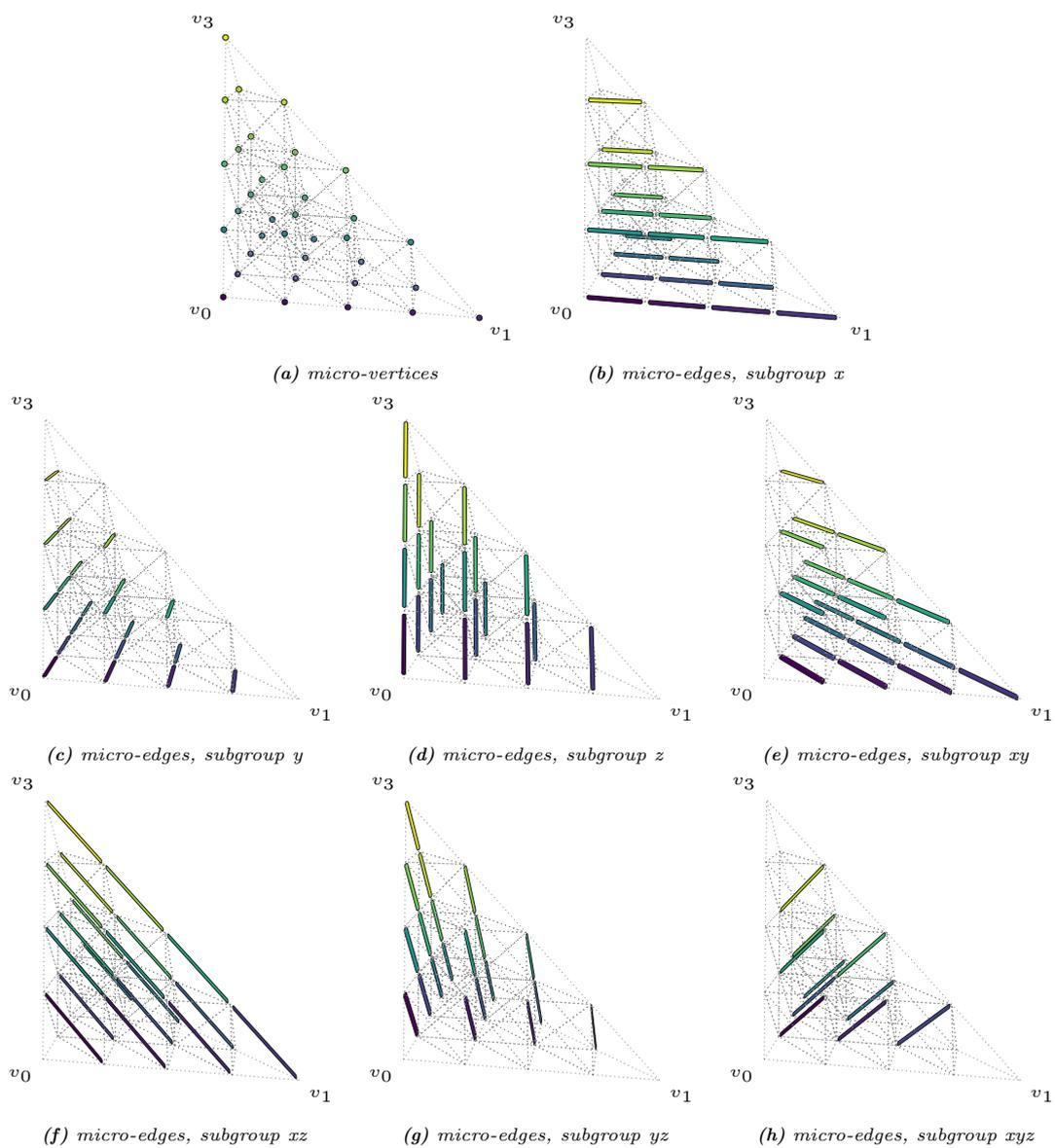

    \footnotesize
    \centering
    \begin{subfigure}{.30\textwidth}
        \resizebox{\linewidth}{!}{\input{figures/micro/micro_vertices_width_5}}
        \caption{micro-vertices}
        \label{fig:appendix-mv}
    \end{subfigure}
    \begin{subfigure}{.30\textwidth}
        \resizebox{\linewidth}{!}{\input{figures/micro/micro_edge_types_width_5_cell_type_EdgeOrientation.X}}
        \caption{micro-edges, subgroup x}
        \label{fig:appendix-me-x}
    \end{subfigure}

    \begin{subfigure}{.30\textwidth}
        \resizebox{\linewidth}{!}{\input{figures/micro/micro_edge_types_width_5_cell_type_EdgeOrientation.Y}}
        \caption{micro-edges, subgroup y}
        \label{fig:appendix-me-y}
    \end{subfigure}
    \begin{subfigure}{.30\textwidth}
        \resizebox{\linewidth}{!}{\input{figures/micro/micro_edge_types_width_5_cell_type_EdgeOrientation.Z}}
        \caption{micro-edges, subgroup z}
        \label{fig:appendix-me-z}
    \end{subfigure}
    \begin{subfigure}{.30\textwidth}
        \resizebox{\linewidth}{!}{\input{figures/micro/micro_edge_types_width_5_cell_type_EdgeOrientation.XY}}
        \caption{micro-edges, subgroup xy}
        \label{fig:appendix-me-xy}
    \end{subfigure}

    \begin{subfigure}{.30\textwidth}
        \resizebox{\linewidth}{!}{\input{figures/micro/micro_edge_types_width_5_cell_type_EdgeOrientation.XZ}}
        \caption{micro-edges, subgroup xz}
        \label{fig:appendix-me-xz}
    \end{subfigure}
    \begin{subfigure}{.30\textwidth}
        \resizebox{\linewidth}{!}{\input{figures/micro/micro_edge_types_width_5_cell_type_EdgeOrientation.YZ}}
        \caption{micro-edges, subgroup yz}
        \label{fig:appendix-me-yz}
    \end{subfigure}
    \begin{subfigure}{.30\textwidth}
        \resizebox{\linewidth}{!}{\input{figures/micro/micro_edge_types_width_5_cell_type_EdgeOrientation.XYZ}}
        \caption{micro-edges, subgroup xyz}
        \label{fig:appendix-me-xyz}
    \end{subfigure}
    \caption{Micro-vertices, and all 7 types of micro-edges, illustrated on refinement level 2.}
    \label{fig:appendix-hyteg-micro-vertices-edges}
\end{figure}

\vspace*{\fill}

\begin{figure}
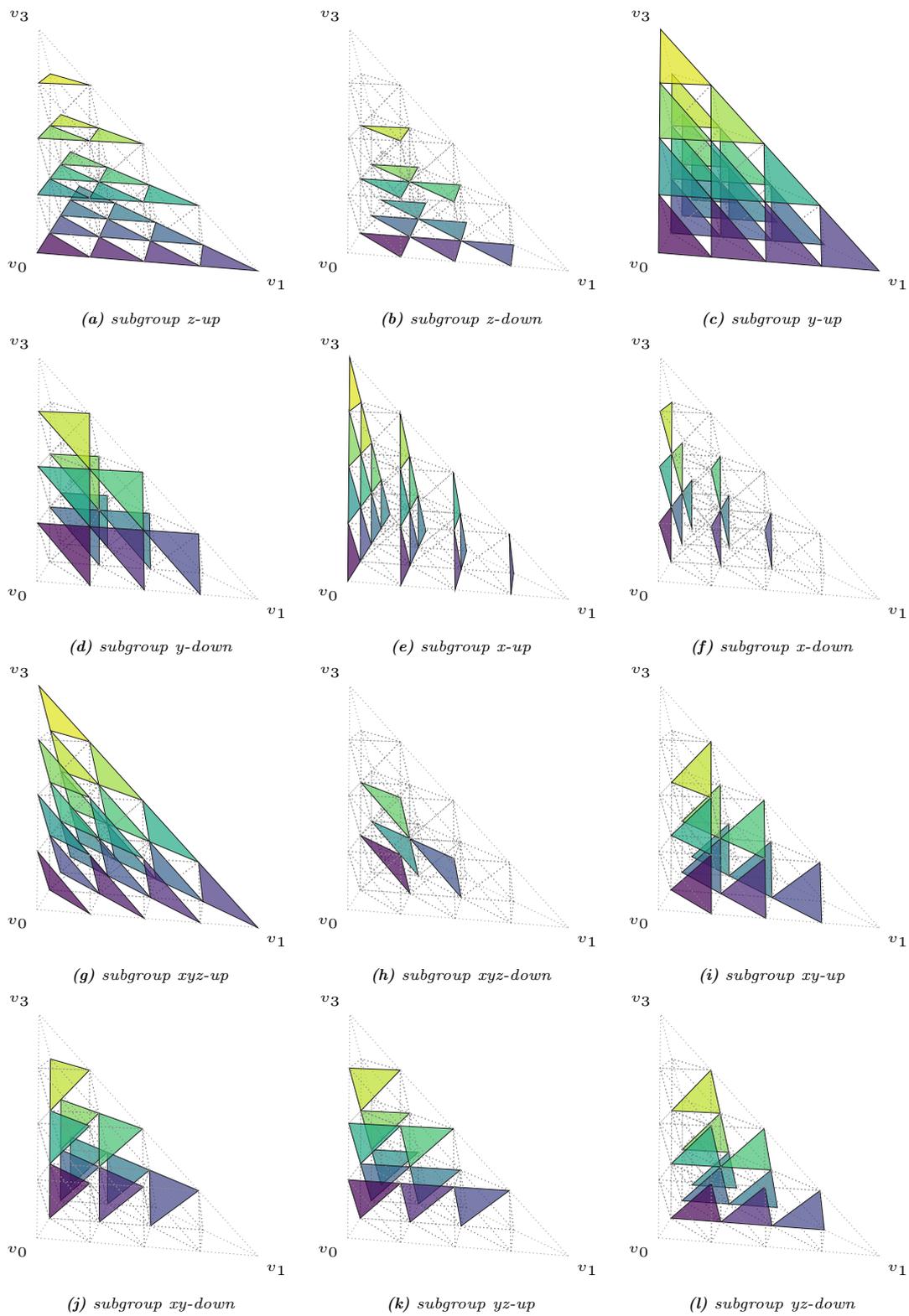

    \footnotesize
    \centering
    \begin{subfigure}{.30\textwidth}
        \resizebox{\linewidth}{!}{\input{figures/micro/micro_face_types_width_5_cell_type_Z_UP}}
        \caption{subgroup z-up}
        \label{fig:appendix-mf-z-up}
    \end{subfigure}
    \begin{subfigure}{.30\textwidth}
        \resizebox{\linewidth}{!}{\input{figures/micro/micro_face_types_width_5_cell_type_Z_DOWN}}
        \caption{subgroup z-down}
        \label{fig:appendix-mf-z-down}
    \end{subfigure}
    \begin{subfigure}{.30\textwidth}
        \resizebox{\linewidth}{!}{\input{figures/micro/micro_face_types_width_5_cell_type_Y_UP}}
        \caption{subgroup y-up}
        \label{fig:appendix-mf-y-up}
    \end{subfigure}

    \begin{subfigure}{.30\textwidth}
        \resizebox{\linewidth}{!}{\input{figures/micro/micro_face_types_width_5_cell_type_Y_DOWN}}
        \caption{subgroup y-down}
        \label{fig:appendix-mf-y-down}
    \end{subfigure}
    \begin{subfigure}{.30\textwidth}
        \resizebox{\linewidth}{!}{\input{figures/micro/micro_face_types_width_5_cell_type_X_UP}}
        \caption{subgroup x-up}
        \label{fig:appendix-mf-x-up}
    \end{subfigure}
    \begin{subfigure}{.30\textwidth}
        \resizebox{\linewidth}{!}{\input{figures/micro/micro_face_types_width_5_cell_type_X_DOWN}}
        \caption{subgroup x-down}
        \label{fig:appendix-mf-x-down}
    \end{subfigure}

    \begin{subfigure}{.30\textwidth}
        \resizebox{\linewidth}{!}{\input{figures/micro/micro_face_types_width_5_cell_type_XYZ_UP}}
        \caption{subgroup xyz-up}
        \label{fig:appendix-mf-xyz-up}
    \end{subfigure}
    \begin{subfigure}{.30\textwidth}
        \resizebox{\linewidth}{!}{\input{figures/micro/micro_face_types_width_5_cell_type_XYZ_DOWN}}
        \caption{subgroup xyz-down}
        \label{fig:appendix-mf-xyz-down}
    \end{subfigure}
    \begin{subfigure}{.30\textwidth}
        \resizebox{\linewidth}{!}{\input{figures/micro/micro_face_types_width_5_cell_type_XY_UP}}
        \caption{subgroup xy-up}
        \label{fig:appendix-mf-xy-up}
    \end{subfigure}

    \begin{subfigure}{.30\textwidth}
        \resizebox{\linewidth}{!}{\input{figures/micro/micro_face_types_width_5_cell_type_XY_DOWN}}
        \caption{subgroup xy-down}
        \label{fig:appendix-mf-xy-down}
    \end{subfigure}
    \begin{subfigure}{.30\textwidth}
        \resizebox{\linewidth}{!}{\input{figures/micro/micro_face_types_width_5_cell_type_YZ_UP}}
        \caption{subgroup yz-up}
        \label{fig:appendix-mf-yz-up}
    \end{subfigure}
    \begin{subfigure}{.30\textwidth}
        \resizebox{\linewidth}{!}{\input{figures/micro/micro_face_types_width_5_cell_type_YZ_DOWN}}
        \caption{subgroup yz-down}
        \label{fig:appendix-mf-yz-down}
    \end{subfigure}

    \caption{All 12 types of micro-faces, illustrated on refinement level 2.}
    \label{fig:appendix-hyteg-micro-faces}
\end{figure}

\begin{figure}
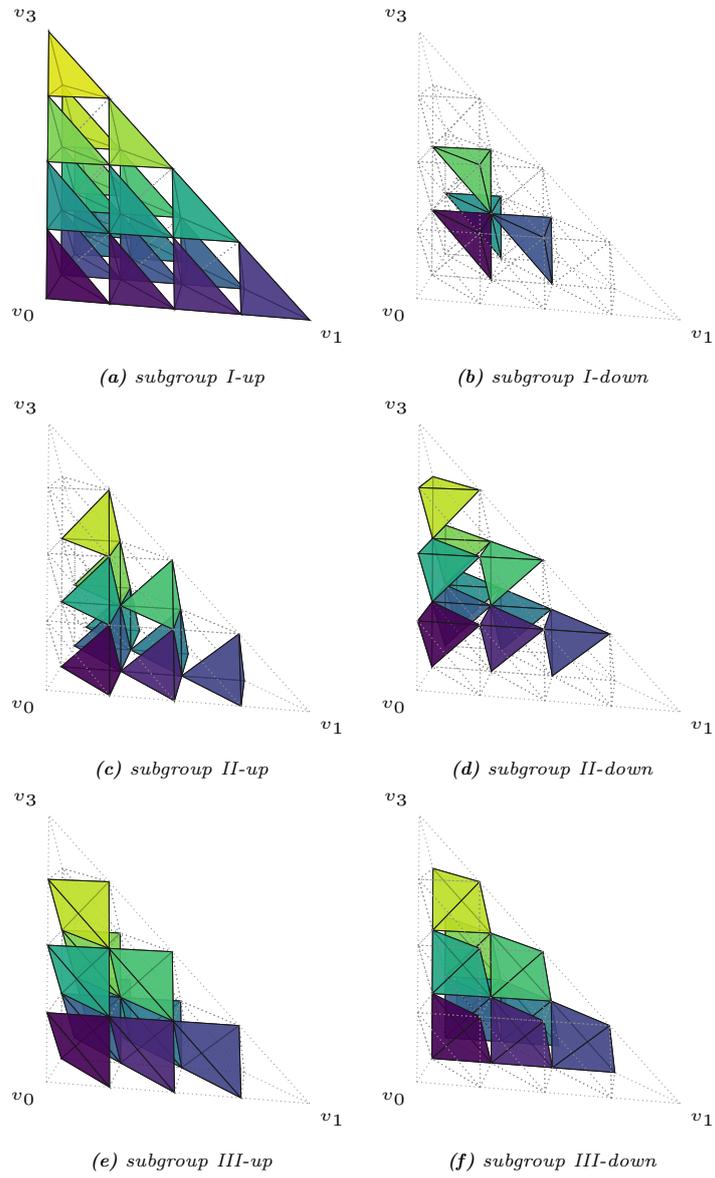

    \footnotesize
    \centering
    \begin{subfigure}{.30\textwidth}
        \resizebox{\linewidth}{!}{\input{figures/micro/micro_cell_types_width_5_cell_type_WHITE_UP}}
        \caption{subgroup I-up}
        \label{fig:appendix-mc-I-up}
    \end{subfigure}
    \begin{subfigure}{.30\textwidth}
        \resizebox{\linewidth}{!}{\input{figures/micro/micro_cell_types_width_5_cell_type_WHITE_DOWN}}
        \caption{subgroup I-down}
        \label{fig:appendix-mc-I-down}
    \end{subfigure}

    \begin{subfigure}{.30\textwidth}
        \resizebox{\linewidth}{!}{\input{figures/micro/micro_cell_types_width_5_cell_type_BLUE_UP}}
        \caption{subgroup II-up}
        \label{fig:appendix-mc-II-up}
    \end{subfigure}
    \begin{subfigure}{.30\textwidth}
        \resizebox{\linewidth}{!}{\input{figures/micro/micro_cell_types_width_5_cell_type_BLUE_DOWN}}
        \caption{subgroup II-down}
        \label{fig:appendix-mc-II-down}
    \end{subfigure}

    \begin{subfigure}{.30\textwidth}
        \resizebox{\linewidth}{!}{\input{figures/micro/micro_cell_types_width_5_cell_type_GREEN_UP}}
        \caption{subgroup III-up}
        \label{fig:appendix-mc-III-up}
    \end{subfigure}
    \begin{subfigure}{.30\textwidth}
        \resizebox{\linewidth}{!}{\input{figures/micro/micro_cell_types_width_5_cell_type_GREEN_DOWN}}
        \caption{subgroup III-down}
        \label{fig:appendix-mc-III-down}
    \end{subfigure}
    \caption{All 6 types of micro-cells, illustrated on refinement level 2.}
    \label{fig:appendix-hyteg-micro-cells}
\end{figure}

\end{document}